\documentclass[twocolumn,aps,floatfix,superscriptaddress]{revtex4-2}
\pdfoutput=1 
\usepackage{amsmath,amssymb,eucal,graphicx,amsbsy}
\usepackage{epsfig}

\usepackage[colorlinks=true, urlcolor=blue, anchorcolor=blue, citecolor=blue,filecolor=blue,linkcolor=blue,menucolor=blue]{hyperref}
\usepackage{soul}
\usepackage{comment}

\begin{document}
\author{P. L. Krapivsky}
\affiliation{Department of Physics, Boston University, Boston, MA 02215, USA}
\affiliation{Santa Fe Institute, Santa Fe, New Mexico 87501, USA}
\author{S. A. Matveev}
\affiliation{Faculty of Computational Mathematics and Cybernetics, Lomonosov Moscow State University, Moscow, 119991, Russia}
\affiliation{Marchuk Institute of Numerical Mathematics, Russian Academy of Sciences, Moscow, 119333, Russia}
\title{Gelation in input-driven aggregation}
\begin{abstract}
We investigate irreversible aggregation processes driven by a source of small mass clusters. In the spatially homogeneous situation, a well-mixed system is consists of clusters of various masses whose concentrations evolve according to an infinite system of nonlinear ordinary differential equations. We focus on the cluster mass distribution in the long time limit. An input-driven aggregation with rates proportional to the product of merging partners undergoes a percolation transition. We examine this process analytically and numerically. There are two theoretical schemes and two natural ways of numerical integration on the level of a truncated system with a finite number of equations. After the percolation transition, the behavior depends on the adopted approach: The giant component quickly engulfs the entire system (Flory approach), or a non-trivial stationary mass distribution emerges (Stockmayer approach). We also outline generalization to ternary aggregation. 
\end{abstract}
\maketitle

\section{Introduction}

Aggregation processes underlie numerous physical and chemical phenomena ranging from micro- to macro-scale \cite{Flory53,book,Leyvraz03}. Examples include blood clotting \cite{anand2006model}, formation of aerosol particles in the atmosphere \cite{voloshchuk1977kinetics,dust93,aloyan1997transport,Friedlander,lushnikov2010introduction}, molecular beam epitaxy \cite{kaganer2016nucleation, hinnemann2003epitaxial}), astrophysical systems \cite{Esposito,Guettler10,Blum11,brilliantov2015size}, etc. 

We assume that each cluster participating in aggregation consists of a finite number of elementary blocks (monomers). This description is appropriate in many situations, e.g., in polymer science \cite{Flory53}, and it provides an accurate description of the most interesting large mass behavior in systems with continuous mass distribution. 

Aggregation proceeds through merging events. Binary merging events dominate in many applications, particularly in dilute systems. Symbolically, the binary aggregation process is represented by the reaction scheme 
\begin{equation}
\label{eq:agg_react}
[i]+[j]  \xrightarrow[]{K_{i,j}} [i+j]
\end{equation}
with aggregation rates $K_{i,j} = K_{j,i} \geq 0$. The set of kinetic coefficients $K_{i,j}$ is called an aggregation kernel. We consider spatially homogeneous systems and study concentrations $c_s(t)$ of the particles of size $s$ per unit volume. The kinetic description of such a system is provided by Smoluchowski equations \cite{smoluchowski1916drei, lushnikov1973evolution, galkin2001}. These equations and generalizations accounting for shattering (possible complete fragmentation into monomers in binary collisions) \cite{matveev2017oscillations,Wendy21,Wendy23}, spontaneous binary fragmentation \cite{bodrova2019kinetic}, exchange-driven reactions \cite{Ke02,PK_exchange03,PK_exchange15,pego2020temporal}, etc. \cite{hayakawa87,esenturk2020role,osinsky2022anomalous,budzinskiy2021hopf} form an active research area. 

We study aggregation processes driven by a constant source of small mass clusters. The state of the system before we turn on the source and the details of the source play a minor role. For concreteness, we assume that the system is initially empty and limit ourselves to the mono-disperse source. We consider the most natural source of monomers, so each cluster is composed of an integer number $s$ of monomers. The reaction scheme $\emptyset \to [1]$ representing the source emphasizes that the source is external, i.e., decoupled from the merging process. Equations 
\begin{eqnarray}
\label{Smol}
\frac{d c_s}{dt} = \frac{1}{2} \sum_{i+j=s} K_{i,j} c_i c_j - c_s \sum_{j=1}^{\infty} K_{s,j} c_j + \delta_{s, 1}
\end{eqnarray}
describe the evolution of concentrations $c_s(t)$ of clusters of mass $s$ in the binary aggregation process \eqref{eq:agg_react} driven by constant input of monomers. 

The input-driven binary aggregation has already been investigated both analytically \cite{hayakawa87} and numerically \cite{esenturk2020role}. Among interesting behaviors, we mention scaling solutions \cite{Leyvraz03}, non-decaying oscillations \cite{ball2012collective} (seemingly arising via Hopf bifurcations), oscillating stationary particle size distributions \cite{matveev2020oscillating} and instantaneous gelation \cite{ball2011instantaneous}. 

In the long time limit, concentrations of clusters in input-driven aggregation often become stationary. In contrast, concentrations evolve ad infinitum for pure aggregation processes, and hence input-driven aggregation is often more amenable to analysis than non-driven aggregation. For a broad family of aggregation kernels, the stationary concentrations have a power-law tail, $c_s \sim s^{-\alpha}$ with $\alpha<2$,  see \cite{book,Leyvraz03,Saslaw,White82,hayakawa87,Colm12,fortin2023stability}. In the case of mass-independent rates, $K_{i,j}=\text{const}$, an exact solution is known (see Sec.~\ref{sec:23}), and the tail exponent is $\alpha=3/2$.  

Ternary aggregation, symbolically
\begin{equation}
[i]+[j]+[k]  \xrightarrow[]{K_{i,j,k}} [i+j+k]
\end{equation}
has been also investigated \cite{krapivsky1991aggregation,jiang1994scaling,krapivsky1994diffusion,oshanin1995smoluchowski,brener2014model} both on the mean-field level based on equations like \eqref{Smol} and in low spatial dimensions, particularly in one dimension, where the diffusion-controlled aggregation process is not accounted by the mean-field equations. As in the binary case, the reaction rates $K_{i,j,k}$ are non-negative and symmetric in masses $i,j,k$ of merging aggregates. Generalizations to $n-$body aggregation processes are also possible \cite{krapivsky1991aggregation, jiang1994scaling}. Models with mass-independent rates, $K_{i_1,\ldots,i_n} = 1$, and with rates $K_{i_1,\ldots,i_n} = i_1+\ldots+i_n$ equal to the sum of masses of the reactants, are exactly solvable. The instantaneous gelation is also possible for many-body aggregation problems \cite{jiang1996instantaneous}. In Sec.~\ref{sec:23}, we analyze steady-state solutions for the process with binary and ternary aggregation, both mass-independent.

For some kernels, the input-driven binary aggregation does not admit stationary concentrations. The simplest example is the binary aggregation with product rates $K_{i,j} = ij$ which we analyze in Sec.~\ref{sec:bin_prod}. This process undergoes a gelation transition. Without a source, the model with kernel $K_{i,j} = ij$ is exactly solvable. It goes through the gelation catastrophe \cite{lushnikov1983analytic} due to the emergence of the giant component rapidly engulfing the whole system. The model is essentially identical to Erd\H{os}--R\'{e}nyi random graphs \cite{ER60, Janson93}. The source-driven model with product rates is more challenging for analytical and numerical investigation. In contrast to the model without source, the concentrations $c_s(t)$ do not admit an explicit solution, but we derive an exact parametric representation of the generating function associated with $c_s(t)$. In Sec.~\ref{sec:tern_prod}, we study the pure ternary aggregation with product kernel $K_{i,j,k}= ijk$. 

Analytical solutions are crucial for tests of the numerical methods. Indeed, one must solve a large number of nonlinear kinetic equations and hopefully get faithful conclusions about an infinite system, or at least a system with so many cluster species that the numerical integration is beyond the limit of what is currently feasible with the best available methods. 

Our numerical experiments (Sec.~\ref{sec:numerical}) rely on the second-order Runge-Kutta numerical time-integration method and utilize an approach based on the low-rank matrix structure of the kinetic coefficients for efficient evaluation of the right-hand side. This approach allows us to perform numerical integration of up to $N=2^{20}$ equations within a modest time on a single workstation.

Numerical investigations of gelling systems require efficient algorithms and a careful choice of the target differential equations. One can only perform a numerical integration of a truncated system (i.e., account for a finite number $N$ cluster species).  The convergence to the infinite system holds only before the gelation transition. An advantage of the aggregation process with $K_{i,j} = ij$ is that the loss term in \eqref{Smol} can be replaced by $sc_s M(t)$. We can then rely on the exact expression for mass concentration, $M(t)=t$ in the case of the source of monomers of the unit strength used in \eqref{Smol}. A  truncated system with such carefully chosen loss term accounts for merging of finite clusters with gel. If instead $M(t)$ we use the sum only over first $N$ cluster species, very different behaviors emerge as we demonstrate in Sec.~\ref{sec:numerical}. Amusingly, this natural numerical implementation is essentially equivalent to the Stockmayer treatment of gelation. In gelling systems without input, both standard (Flory) and Stockmayer approaches predict that $c_s(t)$ decay to zero for every fixed $s$ and $t\to\infty$. In gelling systems with input, concentrations also decay to zero in the realm of the Flory approach, while the steady state is reached in the realm of the Stockmayer approach. We determine the steady state analytically (Secs.~\ref{subsec:SA} and \ref{subsec:SA-3}) and confirm our analytical predictions numerically.

\section{Input-driven aggregation with mass-independent rates}
\label{sec:23}

Input-driven aggregation processes in which the steady states emerges are significantly more tractable than standard aggregation processes with never-ending evolution. As a historical illustration, we recall that the aggregation framework based on an infinite set of coupled differential equations \eqref{Smol} without source was introduced by Smoluchowski in the context of Brownian coagulation \cite{Smol17}.  Smoluchowski argued that the appropriate reactions rates for Brownian coagulation in three dimensions are \cite{Smol17,Chandra43}
\begin{eqnarray}
\label{Brown}
K_{i,j} &=& \left(i^{1/3}+j^{1/3}\right)\left(i^{-1/3}+j^{-1/3}\right) \nonumber\\
          &=&  \left(\frac{i}{j}\right)^{1/3}+\left(\frac{j}{i}\right)^{1/3}+2
\end{eqnarray}
Smoluchowski equations with Brownian kernel \eqref{Brown} have never been solved analytically. For the input-driven Brownian coagulation, the emerging steady state admits an analytical description \cite{Colm12}. In particular, the asymptotic decay of the stationary concentration is the same \cite{Colm12} as for the aggregation with mass-independent reaction rates, $c_s\simeq Cs^{-3/2}$, and only the amplitude $C=\sqrt{5/23}(4\pi)^{-1/2}$ is different from the amplitude $C=(4\pi)^{-1/2}$ in the model with mass-independent rates $K_{ij}=2$, see \eqref{ck-asymp}. 

The simplest input-driven binary aggregation with mass-independent merging rates has been studied in the past \cite{Saslaw,White82}. This model still provides the best illustration of the emergence of stationary mass distribution in the input-driven aggregation which first appears controversial as the mass density grows with time ad infinitum. In this section we re-derive the stationary mass distribution, recall why it does not contradict the unlimited mass growth, and then show how to generalize to ternary aggregation with mass-independent merging rates, and to aggregation with both binary and ternary merging events. 

\subsection{Binary aggregation with input}
\label{subsec:binary}

The binary aggregation process with mass-independent merging rates (we choose $K_{i,j} \equiv 2$) and a source of monomers is described by the rate equations
\begin{equation}
\label{ckt}
\frac{d c_s}{dt}=\sum_{i+j=s}c_ic_j-2c_s c +\delta_{s,1}.
\end{equation}
Here $c(t)=\sum_{j\geq 1} c_j(t)$ is the concentration of clusters, $c_1(t)$ is the concentration of monomers and $c_s(t)$ is the concentration of clusters of ``mass'' $s$, that is, clusters composed of $s$ monomers.  We have set the strength of the source of monomers and the rate of merging to unity --- this can always be achieved by appropriate rescaling of the concentrations and time.

\begin{figure}[ht]
    \includegraphics[scale=0.6]{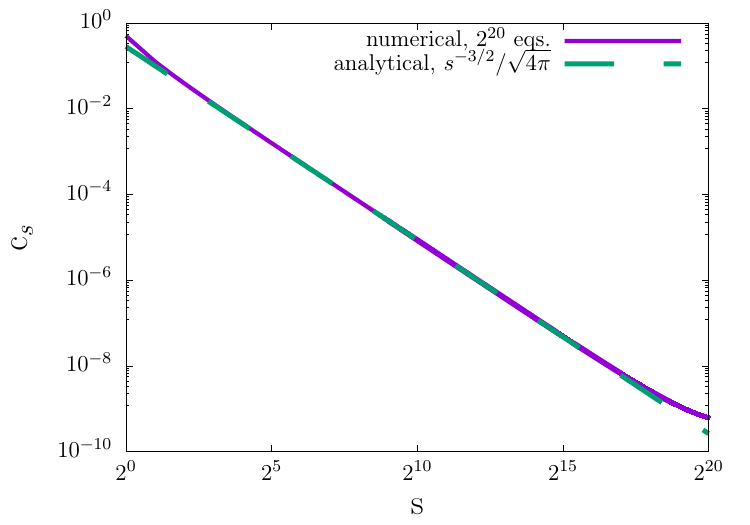} 
\caption{Analytical and numerical results for the mass distributions in input-driven binary aggregation with mass-independent rates 
governed by Eqs.~\eqref{Smol}.}
    \label{fig:pure_bin}
\end{figure}

In the long time limit,  the system reaches a steady state which is universal and well-known \cite{hayakawa87,book,Leyvraz03}, namely independent on the initial condition
\begin{equation}
\label{ck-sol}
c_s=\frac{1}{\sqrt{4\pi}}\,\,
\frac{\Gamma(s-\frac{1}{2})}{\Gamma(s+1)}.
\end{equation}
We are mostly interested in the steady state, so to make formulas less cluttered we write $c_s$ instead of $c_s(\infty)$. The concentrations decay algebraically, viz.
\begin{equation}
\label{ck-asymp}
c_s\simeq \frac{1}{\sqrt{4\pi}}\,\,s^{-3/2}
\end{equation}
when $s\gg 1$. This asymptotic is well supported by numerical integration of a finite number $N\gg 1$ equations when $s\ll N$, see Fig.~\ref{fig:pure_bin}. 

In order to derive \eqref{ck-sol} we use the generating function 
\begin{equation}
\label{Cz}
\mathcal{C}(z)=\sum_{s\geq 1} c_s\,z^s.
\end{equation}
In the steady state, we use \eqref{Cz} and transform Eqs.~\eqref{ckt} into a quadratic equation $0={\cal C}^2-2{\cal C}+z$ for the generating function, from which ${\cal C}(z)=1-\sqrt{1-z}$. Expanding ${\cal C}(z)$ we arrive at \eqref{ck-sol}. Let us now briefly discuss the approach to the steady state \eqref{ck-sol}. The results now depend on the initial condition, and we consider the simplest situation of an initially empty system. The total cluster concentration satisfies 
\begin{subequations}
\begin{equation}
\label{ct-2}
\frac{d c}{dt}=-c^2+1,
\end{equation}
so for initially empty system 
\begin{equation}
\label{ct-sol}
c(t)=\tanh(t).
\end{equation}
\end{subequations}
By inserting \eqref{ct-sol} into the first equation \eqref{ckt} we obtain 
\begin{subequations}
\begin{equation}
\label{c1-eq}
\frac{d c_1}{dt}=-2c_1\tanh(t)+1
\end{equation}
from which
\begin{equation}
\label{c1t-sol}
c_1(t)=\frac{1}{2}\left[\frac{t}{\cosh^2(t)}+\tanh(t)\right].
\end{equation}
\end{subequations}
Using \eqref{ct-sol} we can present Eqs.~(\ref{ckt}) as 
\begin{subequations}
\begin{equation}
\label{cs-eq}
\frac{d c_s}{dt}+2c_s\tanh(t) =\sum_{i+j=s}c_i c_j
\end{equation}
for $s\geq 2$. Multiplying Eq.~\eqref{cs-eq} by $\cosh^2(t)$ we arrive at
\begin{equation}
\frac{d}{dt}\left[c_s(t)\,\cosh^2(t)\right] = \cosh^2(t)\sum_{i+j=s}c_i(t)c_j(t)
\end{equation}
which is integrated to yield 
\begin{equation}
\label{ckt-sol}
c_s(t)=\frac{1}{\cosh^2(t)}\int_0^t dt'\,\cosh^2(t')
\sum_{i+j=s}c_i(t')c_j(t')
\end{equation}
\end{subequations}
allowing (in principle) to obtain all $c_s$ recurrently. Equations \eqref{ct-sol} and \eqref{c1t-sol} show that $c(t)$ and $c_1(t)$ relax exponentially to the steady state values, e.g., $1-c(t)\simeq 2e^{-2t}.$ This holds for other concentrations, but extracting the relaxation and even the final steady-state concentrations from the recursive solution  (\ref{ckt-sol}) is impractical. 

The asymptotic behavior \eqref{ck-asymp} is consistent with the divergence of the mass conservation $\sum sc_s$. Indeed, mass conservation requires $\sum sc_s(t)=t$ at finite time, so the total mass diverges when $t\to\infty$. When $t$ is large, the mass distribution $c_s(t)$ is very close to stationary for sufficiently small masses $s\ll s_*$, while for $s\gg s_*$, the mass distribution is essentially zero. The crossover mass $s_*$ is found from 
\begin{equation}
\label{cross}
t=\sum_{s=1}^\infty sc_s(t)\approx  \sum_{s=1}^{s_*} sc_s\sim \sum_{s=1}^{s_*} s^{-1/2}\sim \sqrt{s_*}
\end{equation}
implying that $s_*\sim t^2$.

\subsection{Ternary aggregation with input}
\label{subsec:ternary}

For aggregation with both two- and three-body merging events and the source of monomers with strength $J$, the governing equations read 
\begin{eqnarray}
\label{long:23}
\frac{d c_s}{dt} &=&  \frac{1}{2} \sum_{i+j=s} K_{i,j} c_i c_j - c_s \sum_{j\geq 1} K_{sj} c_j \nonumber\\ 
& + & \frac{1}{6}\sum_{i+j+k=s} K_{i,j,k} c_i c_j c_k - \frac{c_s}{2}\sum_{i,j\geq 1} K_{s,i,j}  c_i c_j  \nonumber\\ 
&+& J \delta_{s, 1}
\end{eqnarray}
The numerical studies of these equations are complicated due to the higher nonlinearity of the model as well as the complexity of evaluation of the right-hand side, even with a finite number of equations $N$. The straightforward evaluation of the right-hand side for $N$ equations takes $O(N^3)$ operations, making simulations too long even with several thousands of equations. 

This problem requires advanced computational methods. Recently proposed \cite{stefonishin2019tensors, lukashevich2022data} tools allow one to compute the right-hand side within $O(N \log N)$ operations for a special family of kernels having low-rank tensor decomposition in the tensor train \cite{oseledets2011tensor} or canonical polyadic \cite{bro1997parafac} format. Such a decrease in complexity allows one to deal with hundreds of thousands or even millions of kinetic equations using modest computing resources.

For pure ternary aggregation ($K_{i,j}\equiv 0$) with mass-independent merging rates, we set $K_{i,j,k}=6$ and $J=2$ by properly re-scaling the units of concentrations and time. The governing equations become
\begin{equation}
\label{ckt-3}
\frac{d c_s}{dt}=\sum_{i+j+k=s}c_ic_j c_k -3c_s c^2+2\delta_{s,1} .
\end{equation}
The total cluster concentration varies according to 
\begin{equation}
\label{ct-3}
\frac{d c}{dt}=-2c^3+2
\end{equation}
The stationary value is $c=1$ explaining the choice of the strength of the source of monomers. The time-dependent exact (albeit implicit) solution to \eqref{ct-3} subject to $c(0)=0$ can also be found 
\begin{equation}
\label{ct-3-imp}
\ln\frac{\sqrt{1+c+c^2}}{1-c}+ \sqrt{3}\tan^{-1}\left(\frac{1+2c}{\sqrt{3}}\right)=6t+\frac{\pi \sqrt{3}}{6}
\end{equation}
The relaxation to the steady state value is again exponential: $1-c(t) = A e^{-6t}+O(e^{-12 t})$ with $A= \sqrt{3}\,e^\frac{\pi \sqrt{3}}{6}$. 

Stationary concentrations are encapsulated in the generating function 
\begin{equation}
\label{Cz-3}
\sum_{s\geq 1} c_s\,z^s = 2 \sin \frac{\arcsin z}{3} 
\end{equation}
In the steady state, an infinite system \eqref{ckt-3} simplifies to
\begin{equation}
\label{ck-3}
\sum_{i+j+k=s}c_ic_j c_k -3c_s +2\delta_{s,1}=0 
\end{equation}
Using \eqref{Cz} we transform Eqs.~\eqref{ck-3} into a cubic equation $3{\cal C}={\cal C}^3+2z$ for the generating function. One can write a solution in quadratures. It is more convenient, however, to express it through transcendental functions, and this leads to \eqref{Cz-3}. The concentrations $c_s$ with even $s$ equal to zero. The first five non-vanishing concentrations are
\begin{equation*}
\begin{tabular}{ccccc}
$c_1=\frac{2}{3}$, & $c_3= \frac{8}{81}$, & $c_5= \frac{32}{729}$, & $c_7= \frac{512}{19683}$, & $c_9= \frac{28160}{1594323}$
\end{tabular}
\end{equation*}
From the behavior of $\mathcal{C}(z)$ around $z=1$ we find that for odd $s\gg 1$, the asymptotic is 
\begin{equation}
 c_s \simeq \sqrt{\frac{2}{3\pi}}\, s^{-3/2}
\end{equation}
Hence, the decay exponent is the same as in the binary case, cf. Eq.~\eqref{ck-asymp}. 

If both binary and ternary merging events occur and proceed with mass-independent rates
\begin{eqnarray}
\label{ckt-23}
\frac{d c_s}{dt} &=& \lambda \sum_{a+b=s}c_a c_b-2\lambda c_s c \nonumber\\
&+& \sum_{i+j+k=s}c_ic_j c_k -3c_s c^2+(2+\lambda)\delta_{s,1}
\end{eqnarray}
where the strength of the source of monomers was chosen in such way that the stationary total cluster density is $c=1$ as it was in our previous examples of pure binary and pure ternary aggregation. The total cluster density satisfies a rate equation 
\begin{equation}
\label{ct-3lam}
\frac{d c}{dt}=-\lambda c^2 - 2c^3+(2+\lambda)
\end{equation}
admitting an exact but implicit solution as in the case of pure ternary aggregation, cf. Eq.~\eqref{ct-3-imp}. 

\begin{figure}[ht]
\begin{center}
\includegraphics[scale=0.6]{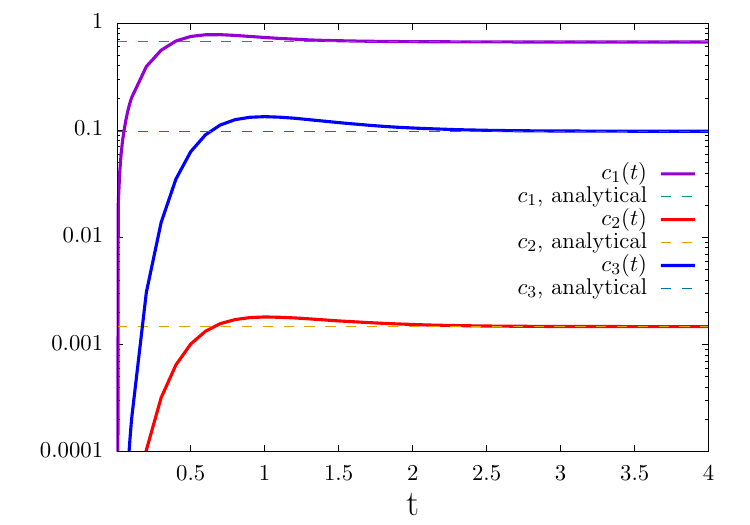}
\caption{Convergence of the concentrations $c_1(t)$, $c_2(t)$, $c_3(t)$ to analytical predictions \eqref{eq:exact_c123} with $\lambda=0.01$. A few hundred equations suffices to obtain very accurate results for the concentrations of light clusters.}
\label{fig:C123_dynamics}
\end{center}
\end{figure}

In the steady state, an infinite system \eqref{ckt-23} simplifies to an infinite set of recurrent equations
\begin{equation*}
\label{ck-23}
\lambda \sum_{a+b=s}c_a c_b+\sum_{i+j+k=s}c_ic_j c_k +(2+\lambda)\delta_{s,1}=(2\lambda+3)c_s.
\end{equation*}
The first three stationary concentrations 
\begin{eqnarray}
\notag && c_1=\frac{\lambda+2}{2\lambda+3}\,, \quad c_2=\frac{\lambda(\lambda+2)^2}{(2\lambda+3)^3}\,, \quad \\
\label{eq:exact_c123}
&& c_3=\frac{(2\lambda^2+2\lambda+3)(\lambda+2)^3}{(2\lambda+3)^5}
\end{eqnarray}
illustrate that results quickly become unwieldy. However, even these few exact expressions are useful for validation of numerical simulations (as we demonstrate in Fig.~\ref{fig:C123_dynamics} and discuss in Sec. \ref{sec:numerical}). 

\begin{figure}[ht]
\begin{center}
    \includegraphics[scale=0.6]{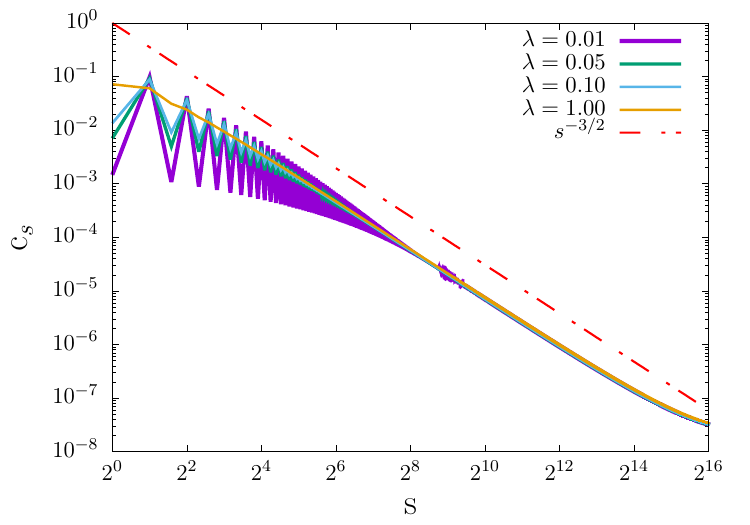}   
    \caption{Numerical results for stationary cluster mass distribution for binary and ternary aggregation for various values of the ratio $\lambda$ 
    characterizing relative strength of the binary aggregation events. The number of equations in numerical integration is $2^{16}$.} 
    \label{fig:Validation_bin_tern}
\end{center}
\end{figure}

The generating function is a root of a cubic equation
\begin{equation}
\label{GF:23}
(2\lambda+3) \mathcal{C} =\lambda \mathcal{C}^2+ \mathcal{C}^3 + (2+\lambda)z.
\end{equation}
One can express $\mathcal{C}(z)$ in quadratures, but extracting explicit general results from such a cumbersome solution seems impossible. The large $s$ behavior is easy to deduce from the asymptotic behavior of $\mathcal{C}(z)$ near $z=1$. Equation \eqref{GF:23} yields 
\begin{equation*}
1-\mathcal{C}\simeq \sqrt{\frac{2+\lambda}{3+\lambda}}\,\, \sqrt{1-z}
\end{equation*}
from which
\begin{equation}
\label{ck-23_final}
 c_s \simeq \sqrt{\frac{2+\lambda}{4\pi(3+\lambda)}}\,\, s^{-3/2}
\end{equation}
when $s\gg 1$. Numerical integration of a finite number $N\gg 1$ equations agree with the asymptotic \eqref{ck-23_final} when $s\ll N$, see Fig.~\ref{fig:Validation_bin_tern}. 

\begin{figure}[ht]
    \includegraphics[scale=0.6]{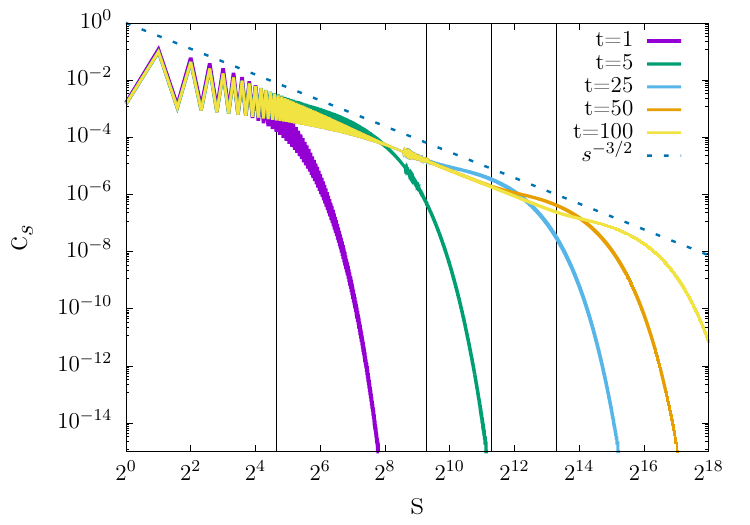}
\caption{Numerical solutions of equations \eqref{ckt-23} with $\lambda = 0.01$ approaches to the analytical prediction \eqref{ck-23_final} for the 
steady state.  The vertical lines correspond to the estimate $s_*= t^2$  of the crossover mass for $t=5$, $25$, $50$, $100$. The number of equations in numerical integration is $2^{18}$.}
    \label{fig:general_dynamics}
\end{figure}

The estimate \eqref{cross} of the crossover time is also supported by numerical integration, see Fig.~\ref{fig:general_dynamics}. 

\section{Binary aggregation with product kernel}
\label{sec:bin_prod}

Input does not necessarily drive an aggregation process to the steady state. We illustrate this assertion for the process with product rates, $K_{ij}=ij$, arguably the most famous aggregation process capturing gelation and percolation phenomena \cite{Flory41,Flory41b, Flory41c, Stockmayer43} that remains an active research subject \cite{Lushnikov04,Lushnikov05,BN-RG,Leyvraz22}. This aggregation process is equivalent to evolving random graphs (also known as Erd\H{os}--R\'{e}nyi random graphs \cite{ER60,Janson93}). Aggregation processes with rates proportional to the product of masses of reactants are popular throughout several branches of sciences, from mathematics and computer science  \cite{Bollobas,Flajolet,Newman-book,Hofstad,Frieze,chatterjee16} to polymer physics and chemistry \cite{Flory53,Ziff80,tanaka1994,tanaka2011book}. 

The influence of input on the aggregation process with reaction rates $K_{ij}=ij$ is intriguing as we show below. 

\subsection{Basic results}

The governing equations read
\begin{equation}
\label{prod}
\frac{d c_s}{dt}=\frac{1}{2}\sum_{i+j=s}ijc_ic_j-t sc_s+\delta_{s,1}
\end{equation}
Initial conditions affect only the earlier behavior. Hence, as before, we focus on the most clean situation of initially empty system, $c_j(0)=0$. With this initial condition
\begin{equation}
\label{mass}
\sum_{j\geq 1}jc_j(t)=t
\end{equation}
explaining the loss term in \eqref{prod}. 

Using \eqref{prod} we deduce the governing equation
\begin{equation}
\label{M2:prod}
\frac{d M_2}{dt}=M_2^2+1
\end{equation}
for the second moment $M_2 = \sum_{k\geq 1} s^2 c_s$. Solving \eqref{M2:prod} subject to $M_2(0)=0$ yields
\begin{equation}
\label{M2:sol}
M_2(t)=\tan t
\end{equation}
The second moment diverges at time 
\begin{equation}
\label{tg:2}
t_g=\frac{\pi}{2}
\end{equation}
when the giant component is born, so \eqref{M2:sol} is applicable only in the pre-gel phase, $t<t_g$. Similarly the third moment $M_3 = \sum_{k\geq 1} s^3 c_s$ satisfies
\begin{equation}
\label{M3:prod}
\frac{d M_2}{dt}=3M_2M_3+1
\end{equation}
whose solution
\begin{equation}
\label{M3:sol}
M_3(t)= \tan t +\tfrac{2}{3}\tan^3 t
\end{equation}
also diverges at $t_g$. Equation \eqref{M3:sol} is also applicable only in the pre-gel phase, $t<t_g$.

Summing Eqs.~\eqref{prod} we deduce the evolution equation 
\begin{equation}
\label{c:prod}
\frac{d c}{dt}=1-\frac{t^2}{2}
\end{equation}
valid when $t<t_g$. Thus
\begin{equation}
\label{M0:sol}
c(t) = \sum_{s\geq 1} c_s(t)  = t-\frac{t^3}{6}
\end{equation}
when $t<t_g$. The zeroth moment $c(t) \equiv M_0(t)$ undergoes a continuous phase transition as we now demonstrate. In the percolating phase, $t>t_g$, the sum $\sum_{j\geq 1}jc_j$ appearing after summation of Eqs.~\eqref{prod} is taken over {\em finite} clusters, and therefore
\begin{equation}
\label{mass-finite}
\sum_{j ~\text{finite}}jc_j(t)=t-g
\end{equation}
where $g$ is the mass of the giant component. Thus instead of \eqref{c:prod} we obtain
\begin{equation}
\label{c:gel-phase}
\frac{d c}{dt}=1-\frac{t^2-g^2}{2}
\end{equation}
Therefore
\begin{equation}
\label{prod-total}
c(t) =
\begin{cases}
t-\frac{t^3}{6}                                                                & t\leq t_g\\
t-\frac{t^3}{6} +\frac{1}{2}\int_{t_g}^t d\tau\,g^2(\tau)   & t>t_g
\end{cases}
\end{equation}

In the long-time limit, every injected monomer is immediately engulfed by the giant component. Hence we deduce $t^2-g^2\to 2$ from \eqref{c:gel-phase} leading, for $t\gg t_g$, to
\begin{equation}
\label{t-g}
t-g\simeq \frac{1}{t}
\end{equation}

One can solve Eqs.~\eqref{prod} recurrently. Solving the rate equation $\dot c_1=1-tc_1$ subject to $c_1(0)=0$ gives 
\begin{subequations}
\begin{equation}
\label{sol:1}
c_1=\int_0^t d\tau\,e^{(\tau^2-t^2)/2}=\sqrt{\frac{\pi}{2}}\,e^{-\frac{t^2}{2}}\,\text{Erfi}\left[\frac{t}{\sqrt{2}}\right]
\end{equation}
where $\text{Erfi}(\cdot)$ is an imaginary error function. In the long-time limit
\begin{equation}
\label{series:1}
c_1=t^{-1}+t^{-3}+3 t^{-5}+15t^{-7}+105t^{-9}+\ldots
\end{equation}
The amplitude in the $t^{-(2n-1)}$ term is $(2n-1)!!$. Therefore, $c_1(t)$ admits an asymptotic expansion
\begin{equation}
\label{asymp:1}
c_1=\sum_{n\geq 1}\frac{(2n-1)!!}{ t^{2n-1}}
\end{equation}
\end{subequations}
Keeping more and more terms in the sum in Eq.~\eqref{asymp:1} provides better and better approximation as the first omitted term gives an estimate of the deviation from the exact result; e.g., \eqref{series:1} is valid up to $O(t^{-11})$. The infinite sum in Eq.~\eqref{asymp:1} diverges as typically happens with asymptotic expansions \cite{Flajolet}. 

We then solve $\dot c_2+2tc_2=c_1^2/2$ and find 
\begin{subequations}
\begin{equation}
\label{sol:2}
c_2 = \frac{\sqrt{\pi}}{2}\,e^{-t^2}\,\text{Erfi}[t]-c_1 + \frac{t}{2}\,c_1^2 
\end{equation}
with $c_1$ given by \eqref{sol:1}. In the long-time limit
\begin{equation}
\label{series:2}
c_2=\tfrac{1}{4}t^{-3}+\tfrac{7}{8} t^{-5}+\tfrac{63}{16}t^{-7}+\tfrac{729}{32}t^{-9}+\ldots
\end{equation}
\end{subequations}

The next equation $\dot c_3+3tc_3=2c_1c_2$ is also solvable:
\begin{subequations}
\begin{equation}
\label{sol:3}
c_3=2e^{-\frac{3t^2}{2}}\int_0^td\tau\,c_1(\tau)c_2(\tau)\,e^{\frac{3\tau^2}{2}}
\end{equation}
In the long-time limit
\begin{equation}
\label{series:3}
c_3=\tfrac{1}{6} t^{-5}+\tfrac{37}{36}t^{-7}+\ldots
\end{equation}
\end{subequations}

The asymptotic behaviors \eqref{series:1} and \eqref{series:2} suggest that in the general case $c_s\simeq s^{-1}A_s t^{-(2s-1)}$ for $t\gg 1$.  Substituting this asymptotic ansatz into Eqs.~\eqref{prod} we arrive at the recurrence 
\begin{equation}
\label{As:rec}
A_s=\frac{1}{2}\sum_{i+j=s}A_iA_j+\delta_{s,1}
\end{equation}
Using the generating function
\begin{equation}
\label{Az}
\mathcal{A}(z)=\sum_{s\geq 1} A_s\,z^s
\end{equation}
we recast the recurrence \eqref{As:rec} into $\mathcal{A}^2-2\mathcal{A}+2z=0$ from which $\mathcal{A}=1-\sqrt{1-2z}$ leading to 
\begin{equation}
\label{Ak-sol}
A_s=\frac{2^{s-1}}{\sqrt{\pi}}\,\,
\frac{\Gamma(s-\frac{1}{2})}{\Gamma(s+1)}
\end{equation}
Therefore
\begin{equation}
\label{ck:prod}
c_s\simeq \frac{2^{s-1}}{\sqrt{\pi}}\,\,
\frac{\Gamma(s-\frac{1}{2})}{s\,\Gamma(s+1)}\,t^{-(2s-1)}
\end{equation}
when $t\gg 1$. If additionally $s\gg 1$, \eqref{ck:prod} becomes
\begin{equation}
\label{ck:prod-asymp}
c_s\simeq \pi^{-1/2} s^{-5/2}\, \frac{2^{s-1}}{t^{2s-1}}
\end{equation} 

Recall that without input the mass of finite clusters vanishes exponentially, namely as $e^{-t}$. In the same system with input the mass of finite clusters decays algebraically, namely as $1/t$ in the long time limit, Eq.~\eqref{t-g}.

\subsection{Generating function technique}

We have established a few exact results in the pre-percolation phase; see \eqref{M2:sol}, \eqref{M3:sol} and \eqref{M0:sol}.  We have also shown how to compute $c_s(t)$ throughout the evolution, $0<t<\infty$, and presented explicit results for small clusters: \eqref{sol:1} and \eqref{sol:2}. All $c_s(t)$ remain infinitely smooth throughout the evolution, yet there is a phase transition at $t=t_g$. The challenge is finding the behavior of $g(t)$ in the percolating phase, $t>t_g$. The recurrent nature of the governing equations \eqref{prod} suggests employing the generating function technique  \cite{Knuth,Flajolet}. 

\subsubsection{Exponential generating function}

Instead of the ordinary generating function \eqref{Cz} it is convenient to use an exponential generating function
\begin{equation}
\label{Cz:exp}
\mathcal{E}(z,t)=\sum_{s\geq 1} s c_s(t)\,e^{s z}
\end{equation}
associated with the sequence $s c_s(t)$. Multiplying Eq.~\eqref{prod} by $s e^{sz}$ and summing over all $s\geq 1$ we reduce an infinite system \eqref{prod} of ordinary differential equations to a single hyperbolic partial differential equation
\begin{equation}
\label{Burgers}
\partial_t \mathcal{F}=\mathcal{F}\partial_z \mathcal{F}+e^z-1, \qquad \mathcal{F}(z,t)\equiv \mathcal{E}(z,t)-t
\end{equation}
The homogeneous part of this equation is the (inviscid) Burgers equation. The general solution of \eqref{Burgers} is found by the method of characteristics 
\begin{subequations}
\begin{align}
\label{Burgers:sol}
&t+\int_{z}^0 \frac{dw}{\sqrt{2(f-e^w+w)}}=\Phi(f)  \\
\label{f:def}
&f=f(z,t)=e^z-z+\tfrac{1}{2}\,\mathcal{F}(z,t)^2
\end{align}
\end{subequations}
To fix $\Phi$, we note that for the initially empty system
\begin{equation}
\mathcal{E}(z,0)=\mathcal{F}(z,0)=0
\end{equation}
Specializing \eqref{Burgers:sol} to $t=0$ yields
\begin{equation}
\label{Phi}
\int_{z}^0 \frac{dw}{\sqrt{2(e^z-z-e^w+w)}}=\Phi(e^z-z)
\end{equation}
which implicitly determines the function $\Phi$.

\subsubsection{Giant component}

Recall that
\begin{equation}
\mathcal{F}(0,t)=\mathcal{E}(0,t)-t=
\begin{cases}
0  &  t<t_g\\
-g  & t>t_g
\end{cases}
\end{equation}
Specializing \eqref{Burgers:sol}--\eqref{f:def} to $z=0$ when $t>t_g$ we obtain
\begin{equation}
\label{tg-Phi}
t = \Phi(1+\tfrac{1}{2}g^2)
\end{equation}

\begin{figure}[ht!]
\includegraphics[width=0.4\textwidth]{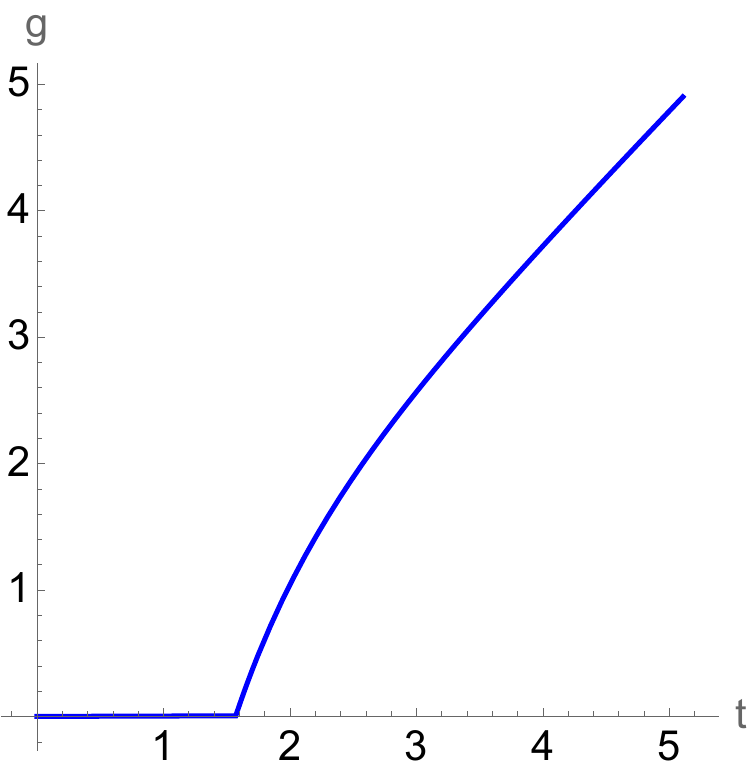}
\caption{The mass $g$ of the giant component vanishes when $t < t_g$. In the percolating phase, $g$ is represented by \eqref{gt}.} 
\label{fig:g}
\end{figure}

Thus, the mass  $g=g(t)$ of the giant component admits a parametric representation 
\begin{equation}
\label{gt}
\begin{split}
& g = \sqrt{2(e^\zeta-\zeta-1)}\\
& t  = \int_{\zeta}^0 \frac{dw}{\sqrt{2(e^\zeta-\zeta-e^w+w)}}
\end{split}
\end{equation}
The mass of the giant component is shown in Fig.~\ref{fig:g}. The asymptotic behaviors admit more explicit descriptions. Massaging the parametric representation \eqref{gt}, one finds such explicit formulae. The asymptotic expansion for the mass of finite clusters in the $t\to\infty$ limit 
\begin{subequations}
\begin{align}
\label{g:inf}
t-g=t^{-1}+\frac{3}{2}\,t^{-3}+\frac{21}{4}\,t^{-5}+\frac{319}{12}\,t^{-7}+\ldots
\end{align}
The asymptotic expansion of the mass of the giant component just above the transition point ($\delta=t-t_g\to +0$)
\begin{align}
\label{g:above}
g=3\delta-\frac{9\pi}{16}\,\delta^2 + \frac{3}{128}(224-96\pi+9\pi^2)\delta^3+\ldots 
\end{align}
\end{subequations}
The asymptotic \eqref{g:inf} is even easier to deduce from \eqref{series:1}, \eqref{series:2}, \eqref{series:3} and $c_4\simeq \frac{5}{32}t^{-7}$. 

The mass $M_1=t-g$ of finite components (see Fig.~\ref{fig:Mass_diff}) reaches the maximum $\text{max}[M_1(t)]=M_1(t_g)=t_g$ at the percolation point; $M_1(t)$ undergoes a continuous phase transition with first derivative exhibiting a jump at the percolation point:
\begin{equation}
\label{M1:jump}
\frac{dM_1}{dt}\Big|_{t_g-0}=1, \qquad \frac{dM_1}{dt}\Big|_{t_g+0}=-2
\end{equation}

\subsubsection{Total cluster concentration}

The total cluster concentration, equivalently the zeroth moment, satisfies $c\equiv M_0\leq M_1$. Hence $c(t)\to 0$ as $t\to\infty$, and integrating \eqref{c:gel-phase} we obtain
\begin{equation}
\label{M0:above}
c(t) = \int_{t}^\infty d\tau\,\frac{\tau^2-g^2(\tau)-2}{2}
\end{equation}
for $t>t_g$. When $t<t_g$, the concentration of clusters is given by \eqref{M0:sol}. However, using \eqref{gt} and \eqref{M0:above} to plot $c(t)$ in the percolating phase is challenging. Since simulations perfectly agree with theoretical predictions, it suffices to plot numerical predictions for $c(t)$. The asymptotic behaviors of $c(t)$ in the percolating phase admit more explicit descriptions. The long-time asymptotic is
\begin{subequations}
\begin{align}
\label{M0:asymp}
c = t^{-1}+\frac{5}{4}\,t^{-3}+\frac{97}{24}\,t^{-5}+\frac{5795}{288}\,t^{-7}+\ldots
\end{align}
Just above the transition point 
\begin{eqnarray}
\label{c:above}
c &=& \frac{\pi}{2}-\frac{\pi^3}{48}+\left(1-\frac{\pi^2}{8}\right)\delta-\frac{\pi}{4}\,\delta^2 + \frac{4}{3}\,\delta^3 \nonumber \\
&-& \frac{27\pi}{64}\,\delta^4+\frac{9(896 - 384 \pi + 45\pi^2)}{2560}\delta^5+\ldots
\end{eqnarray}
\end{subequations}
where $\delta=t-t_g\to +0$. 

\begin{figure}[ht!]
\includegraphics[scale=0.6]{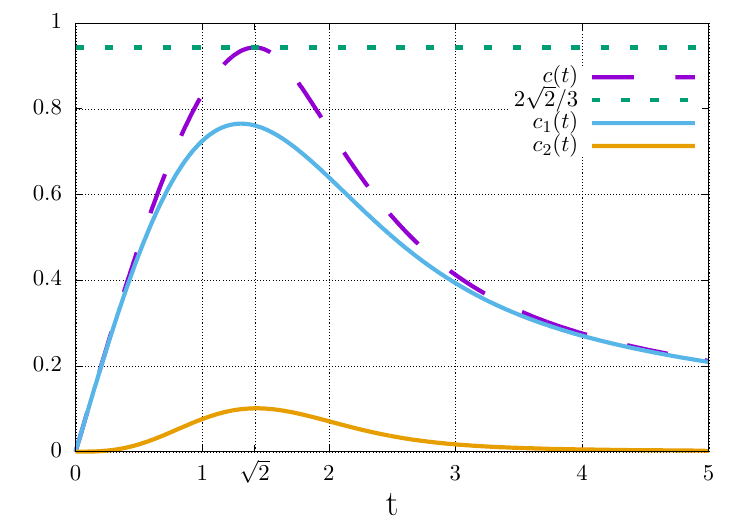}
\caption{Top to bottom: Total cluster concentration $c(t)$, the concentration of monomers $c_1(t)$ given by \eqref{sol:1} and the concentration of dimers $c_2(t)$ given by \eqref{sol:2}. Numerical results for $c(t)$ are in excellent agreement with theoretical predictions, e.g., the maximal value $2\sqrt{2}/3$ is reached at time $t=\sqrt{2}$. } 
\label{fig:c12}
\end{figure}

The total cluster concentration reaches maximum at $t=\sqrt{2}$ where $\text{max}[c(t)]=c\big(\sqrt{2}\big)=2\sqrt{2}/3$, see Fig.~\ref{fig:c12}. 
The maximum occurs before the percolation point; $c(t)$ is a decreasing function of time when $\sqrt{2}<t\leq t_g$ and in the entire percolating phase $t>t_g$.  At the percolation point, $c(t)$ undergoes a continuous phase transition. More precisely, $c$ together with first and second derivatives are continuous at $t=t_g$ and equal to
\begin{subequations}
\begin{equation}
\label{M0:012}
c=\frac{\pi}{2}-\frac{\pi^3}{48}\,, \quad \frac{d c}{dt}=1-\frac{\pi^2}{8}\,, \quad  \frac{d^2 c}{dt^2}=-\frac{\pi}{2}\\
\end{equation}
while the third derivative exhibits a jump:
\begin{align}
\label{M0:jump}
\frac{d^3 c}{dt^3}\Big|_{t_g-0}=-1, \qquad \frac{d^3 c}{dt^3}\Big|_{t_g+0}=8
\end{align}
\end{subequations}

\subsubsection{Moments $M_n$ with $n\geq 2$}

The moments $M_0$ and $M_1$ remain finite throughout the evolution. The moments $M_n$ with $n\geq 2$ diverge at the percolation point. These moments can still be defined in the percolating phase if the sum is taken over finite components:
\begin{equation}
\label{Mn:finite}
M_n(t) = \sum_{j ~\text{finite}}j^n c_j(t)
\end{equation}
All these moments with integer $n\geq 2$ are encapsulated in the derivatives of the generating function:
\begin{equation}
\label{Mn}
M_n=(\partial_z)^{n-1} \mathcal{F}\big|_{z=0}
\end{equation}
This formula applies to the entire time range. 

Specializing \eqref{Burgers} to $z=0$ and recalling that
\begin{equation}
\mathcal{F}\big|_{z=0}=-g, \qquad \partial_z \mathcal{F}\big|_{z=0}=M_2
\end{equation}
we deduce a neat formula expressing the second moment via the mass of the giant component:
\begin{equation}
\label{M2:finite}
M_2 = g^{-1}\,\frac{dg}{dt}
\end{equation}
As a consistency check, one can use \eqref{series:1}, \eqref{series:2}, \eqref{series:3}, etc. to deduce the asymptotic expansion of $M_2$. Using the asymptotic expansion \eqref{g:inf}, one obtains the asymptotic expansion of the right-hand side of Eq.~\eqref{M2:finite}. The expansions coincide in all orders. 

Higher moments can be similarly extracted, e.g., the third moment can be obtained by applying $ \partial_z$ to \eqref{Burgers} and then specializing to $z=0$. Equivalently, we can begin with the evolution equation for the second moment
\begin{equation}
\label{M2:finite-eq}
\frac{d M_2}{dt}=1+M_2^2 - gM_3
\end{equation}
The extra term compared to Eq.~\eqref{M2:prod} arises since we must be more careful in the percolating phase and include $M_3(\sum_{s\geq 1}s c_s-t)=-gM_3$; when $t<t_g$, this term vanishes and hence does not appear in Eq.~\eqref{M2:prod}. The hierarchical nature of \eqref{M2:finite-eq} does not allow one to determine $M_2$ from \eqref{M2:finite-eq}. However, we already established $M_2$ using a different approach, and hence we can now use \eqref{M2:finite-eq} to determine the third moment:
\begin{equation}
\label{M3:finite}
M_3 = g^{-1}\left[1+M_2^2-\frac{dM_2}{dt}\right]
\end{equation}
Since $M_2$ given by \eqref{M2:finite} is expressed via $g$, Eq.~\eqref{M3:finite} expresses $M_3$ via $g$. 

The same strategy works for higher moments. Using \eqref{prod} we deduce the evolution equation 
\begin{equation}
\label{M3:finite-eq}
\frac{d M_3}{dt}=1+3M_2M_3 - gM_4
\end{equation}
from which
\begin{equation}
\label{M4:finite}
M_4 = g^{-1}\left[1+3M_2M_3-\frac{dM_3}{dt}\right]
\end{equation}
effectively expressing $M_4$ via $g$. Continuing, one writes the evolution equation for the forth moment from which one extracts the fifth moment 
\begin{equation}
\label{M5:finite}
M_5 = g^{-1}\left[1+4M_2M_4+3M_3^2-\frac{dM_4}{dt}\right]
\end{equation}
This equation in conjunction with previous results for the moments $M_2, M_3, M_4$ effectively expresses $M_5$ via $g$.

\subsubsection{Cluster mass distribution}

At the percolation point, Eqs.~\eqref{Burgers:sol}--\eqref{f:def} become
\begin{subequations}
\begin{align}
\label{Burgers:crit}
&\frac{\pi}{2}+\int_{z}^0 \frac{dw}{\sqrt{2(f_*-e^w+w)}}=\Phi(f_*)  \\
\label{f:crit}
&f_*=e^z-z+\tfrac{1}{2}\,\mathcal{F}_*^2, \qquad \mathcal{F}_*=\mathcal{F}(z,t_g)
\end{align}
\end{subequations}
and the cluster mass distribution has an algebraic tail
\begin{equation}
\label{cs:prod-crit}
c_s(t_g)\simeq C\,s^{-5/2}\qquad\text{as}\quad s\to\infty
\end{equation}
By inserting this asymptotic into the definition of the exponential generating function we deduce the asymptotic 
\begin{equation}
\label{F:crit}
\mathcal{F}_* = -B\sqrt{-z}+O(z), \qquad B=C\Gamma\big(-\tfrac{1}{2}\big)
\end{equation}
Combining \eqref{f:crit} and \eqref{F:crit} we obtain
\begin{equation}
\label{f:crit-exp}
f_* = 1 - \tfrac{1}{2}z B^2+ O(z^2)
\end{equation}

By inserting \eqref{f:crit-exp} into the left-hand side (LHS) of \eqref{Burgers:crit} we obtain
\begin{equation}
\label{LHS:crit}
\text{LHS} = \frac{\pi}{2}-B^{-1}\sqrt{-z}+O(z)
\end{equation}
Using \eqref{Phi} and expanding near the origin one gets
\begin{equation}
\label{Phi:exp}
\Phi\left(1+\frac{\zeta^2}{2}+\ldots\right)=\frac{\pi}{2}-\frac{\zeta}{3}+\ldots
\end{equation}
which in conjunction with  \eqref{f:crit-exp} leads to the expansion
of the right-hand side (RHS) of \eqref{Burgers:crit}
\begin{equation}
\label{RHS:crit}
\text{RHS} = \frac{\pi}{2}-\frac{B}{3}\,\sqrt{-z}+O(z).
\end{equation}
Comparing \eqref{LHS:crit} and \eqref{RHS:crit} we fix $B=-\sqrt{3}$. Thus we confirm \eqref{cs:prod-crit} and fix the amplitude
\begin{equation}
\label{cs:crit}
c_s\simeq \sqrt{\frac{3}{4\pi}}\,\,s^{-5/2}\qquad\text{as}\quad s\to\infty
\end{equation}

In the percolating phase ($t>t_g$), the tail is exponential with $s^{-5/2}$ algebraic pre-factor 
\begin{equation}
\label{cs:prod-perc}
c_s(t)\simeq C(t)\,s^{-5/2} e^{-s\Lambda(t)}
\end{equation}
To determine the rate $\Lambda(t)$  of the exponential decay and the amplitude $C(t)$ one should expand Eqs.~\eqref{Burgers:sol}--\eqref{f:def} near $z=\Lambda$. Equating the zeroth order terms gives an implicit equation
\begin{subequations}
\begin{align}
&t = \int_0^\Lambda \frac{dw}{\sqrt{2(\lambda-e^w+w)}} +\Phi(\lambda) \\
&\lambda=e^\Lambda-\Lambda+\tfrac{1}{2}\mathcal{F}(\Lambda,t)^2
\end{align}
\end{subequations}
for $\Lambda$. Equating the $\sqrt{\Lambda-z}$ terms one can fix the amplitude $C(t)$ in \eqref{cs:prod-perc}.

\subsection{Stockmayer approach}
\label{subsec:SA}

So far, we have employed an approach originated in the work of Flory \cite{Flory41,Flory41b, Flory41c}. This approach is physically well-motivated as it accounts for merging between finite clusters. Furthermore, the concept of the order parameter crucial for phase transitions is natural within the Flory framework --- the fraction of mass in the giant component (gel) plays the role of the order parameter. The Flory framework is traditionally used in applications \cite{book,Leyvraz03, thamm2003phase}, and the evolution of random graphs \cite{ER60,Janson93,Bollobas,Flajolet,Hofstad,Newman-book,Frieze,BN-RG,lushnikov2015source} is also treated in the Flory framework. 

Almost simultaneously with Flory \cite{Flory41}, Stockmayer \cite{Stockmayer43} proposed a competing theory, see \cite{Ziff80,Leyvraz03} for modern expositions. Merging of finite clusters to gel is ignored by the Stockmayer approach. Both methods give identical results only in the pre-percolating phase, and the percolation transition occurs at the same moment in both frameworks. The concept of gel becomes ill-defined in the Stockmayer framework. One can still define the order parameter as the mass discrepancy, but it loses the natural meaning of the mass of the gel. The Stockmayer approach has several applications, and it also makes sense in some random graph processes \cite{Curien23,K24-SRG,tanaka1994}. A straightforward numerical integration based on truncating an infinite system of equations to a finite system is essentially equivalent to the Stockmayer approach (Sec.~\ref{sec:numerical}). 

We now analyze the aggregation process with product rates and the source of monomers using the approach of Stockmayer. Mathematically, one ought to solve 
\begin{equation}
\label{prod-S}
\frac{d c_s}{dt}=\frac{1}{2}\sum_{i+j=s}ijc_ic_j-M_1(t) sc_s+\delta_{s,1}
\end{equation}
In the pre-percolating phase, the mass of finite clusters is $M_1=t$. In the post-percolating phase, $M_1$ is a priori unknown and should be determined self-consistently. 

Flory and Stockmayer's approaches predict greatly different asymptotic behaviors. Without input, all concentrations vanish when $t=\infty$ and the asymptotic decay laws greatly differ. For instance, if the evolution begins with the mono-disperse initial condition, $c_s(0)=\delta_{s,1}$, the mass of the finite clusters is $M_1 = 1/t$ for all $t\geq t_g=1$ in the realm of the Stockmayer approach \cite{Ziff80,Leyvraz03}; the Flory approach gives $M_1=e^{-t}+te^{-2t}+\ldots$ as $t\to\infty$.

For the input-driven aggregation process with product rates, the concentrations vanish in the realm of the Flory approach, while the Stockmayer approach predicts a non-trivial steady state. Moreover, the steady state is universal, i.e., independent of the initial condition. The stationary solution predicted by the Stockmayer approach is non-trivial. The relaxation to the steady state is quick, so we focus on the stationary solution and disregard the relaxation.

The determination of the stationary mass distribution $c_s$, the cluster concentration $c$ and $\mu=M_1$ is surprisingly easy if we rely on an `experimental' observation following (see Fig.~\ref{fig:Strange_cs}) from numerical integration: $c_s$ decays slower than exponentially with mass $s$. Indeed, in the steady state \eqref{prod-S} simplifies to
\begin{equation}
\label{As-mu:rec}
\mu A_s=\frac{1}{2}\sum_{i+j=s}A_iA_j+\delta_{s,1}
\end{equation}
where $A_s=s c_s$.  Using the generating function \eqref{Az} we recast the recurrence \eqref{As-mu:rec} into $\mathcal{A}^2-2\mu \mathcal{A}+2z=0$ from which $\mathcal{A}=\mu - \sqrt{\mu^2-2z}$. Different behaviors emerge depending on whether $\mu$ is smaller, equal, or larger than $\sqrt{2}$. If $\mu<\sqrt{2}$, the dominating behavior of $A_s$ is the exponential $A_s \propto (2/\mu^2)^s$ growth which is inconsistent with $\mu=\sum_{s\geq 1} A_s$ being finite. If $\mu>\sqrt{2}$, the dominating behavior is the exponential $A_s \propto (2/\mu^2)^s$ decay, which is inconsistent with numerical integration. Therefore
\begin{subequations}
\begin{equation}
\label{M1-S}
\mu=M_1(\infty)=\sqrt{2}
\end{equation}
and using $\mathcal{A}=\sum_{s\geq 1}sc_s z^s =\sqrt{2}\left(1 - \sqrt{1-z}\right)$ and expanding the right-hand side we find stationary concentrations
\begin{equation}
\label{cs-S}
c_s=\frac{1}{\sqrt{2\pi}}\,\,
\frac{\Gamma(s-\frac{1}{2})}{s\Gamma(s+1)}
\end{equation}
The total cluster concentration is found by summing \eqref{cs-S} to give
\begin{equation}
\label{c-S}
c = (1-\ln 2)\sqrt{8}=0.8679108378\ldots
\end{equation}
\end{subequations}

\section{Ternary aggregation with product kernel}
\label{sec:tern_prod}

Here, we consider a ternary aggregation process with reaction rates equal to the product of masses of the merging clusters. This aggregation process is equivalent to evolving random two-dimensional simplicial complexes. Traditionally, one starts with a set with $N$ vertices and creates triangles consisting of (randomly chosen) triplets of vertices. We shall continue to talk about clusters, i.e., maximal connected components, and we characterize each cluster by the number of vertices. This characterization is formally incomplete. However, almost all clusters are topologically similar, namely their Euler characteristic
$\chi=V-E+T$ is $\chi=1$. Here $V$ is the number of vertices, $E$ is the number of edges, $T$ is the number of triangles. (Similarly, in the Erd\H{os}--R\'{e}nyi random graphs almost all clusters are trees with Euler characteristic $\chi=V-E=1$.) 

\subsection{Flory approach}
\label{subsec:FA-3}

For the input-driven pure ternary aggregation process with reaction rates $K_{i,j,k}=ijk$ and the strength of source $J=1$, Eqs.~\eqref{long:23} become  
\begin{equation}
\label{prod3}
\frac{d c_s}{dt}=\frac{1}{6}\sum_{i+j+k=s}ijkc_ic_jc_k  - \frac{1}{2}\,t^2 sc_s+\delta_{s,1}
\end{equation}
In the initially empty system, only clusters with odd mass are formed. 

As in the binary case, we detect the formation of the gel from the divergence of the second moment. Using \eqref{prod3} we deduce 
\begin{equation}
\label{M2:prod3}
\frac{d M_2}{dt}=tM_2^2+1
\end{equation}
which is solved to yield
\begin{equation}
\label{M2:sol3}
M_2(t)=\frac{\text{Bi}(-t)-\sqrt{3}\,\text{Ai}(-t)}{\sqrt{3}\,\text{Ai}'(-t)-\text{Bi}'(-t)}
\end{equation}
where $\text{Ai}$ and $\text{Bi}$ are Airy functions and $f'(x)=df/dx$. The second moment diverges at time 
\begin{equation}
\label{tg:3}
t_g=1.514\,906\,050\ldots
\end{equation}
found from $\sqrt{3}\,\text{Ai}'(-t_g)=\text{Bi}'(-t_g)$, The giant component appears when $t>t_g$, so \eqref{M2:sol3} is applicable only in the pre-percolation phase, $t<t_g$. 

The third moment satisfies
\begin{equation}
\label{M3:prod3}
\frac{d M_3}{dt}=3tM_2M_3+M_2^3+1
\end{equation}
whose solution
\begin{equation}
\label{M3:sol3}
M_3= \int_0^t d\tau\left[M_2^3(\tau)+1\right]e^{3\int_{\tau}^t dt_1 \,t_1 M_2(t_1)}
\end{equation}
also diverges at $t_g$. All moments $M_n(t)$ with $n\geq 2$ diverge at $t_g$. The analytical expressions of $M_n(t)$ with $n\geq 4$ in the pre-percolating phase quickly become unwieldy as $n$ increases. 

The zeroth moment remains finite throughout the evolution. This assertion follows already from $c\equiv M_0<M_1$. In the pre-percolating phase, the zeroth moment obeys
\begin{equation}
\label{ct:3}
\frac{d c}{dt}=1  - \frac{1}{3}\,t^3
\end{equation}
from which
\begin{equation}
\label{ct:sol3}
c = t  - \frac{1}{12}\,t^4
\end{equation}
for $t<t_g$. 
The maximum is again reached before $t_g$, namely at $t_*=\sqrt[3]{3}$ where $c_*=3\sqrt[3]{3}/4$ (see Fig. \ref{fig:ternary_gelation}). In the percolating phase
\begin{equation}
\label{ct:3+}
\frac{d c}{dt}=1  - \frac{1}{3}\,t^3+\frac{3tg^2-g^3}{6}
\end{equation}

Solving Eqs.~\eqref{prod3} recurrently one finds formulae for concentrations valid during the entire evolution:
\begin{subequations}
\begin{align}
\label{sol:1-3}
c_1 &=e^{-t^3/6}\int_0^t d\tau\,e^{\tau^3/6}  \\
\label{sol:3-3}
c_3 &=\frac{1}{6}\,e^{-t^3/2}\int_0^t dt_1\left(\int_0^{t_1} d\tau\,e^{\tau^3/6}\right)^3
\end{align}
\end{subequations}
etc. The large time expansions are
\begin{subequations}
\label{1357}
\begin{align}
\label{asymp:1-3}
c_1 &=\frac{2}{t^2}+\frac{8}{t^5}+\frac{80}{t^8}+\frac{1280}{t^{11}}+\frac{28160}{t^{14}}+\ldots  \\
\label{asymp:3-3}
c_3 &=\tfrac{8}{9}t^{-8}+\tfrac{416}{27}t^{-11}+\tfrac{21248}{81}t^{-14}+\ldots \\
\label{asymp:5-3}
c_5 &=\tfrac{32}{15}t^{-14}+\tfrac{14848}{225}t^{-17}+\tfrac{5328896}{3375}t^{-20}+\ldots \\
\label{asymp:7-3}
c_7 &=\tfrac{512}{81}t^{-20}+\tfrac{2399744}{8505}t^{-23}+\ldots 
\end{align}
\end{subequations}
etc. The asymptotic behaviors \eqref{1357} suggest that in the general case $c_s\simeq s^{-1}A_s t^{-(3s-1)}$ for $t\gg 1$.  Substituting this asymptotic ansatz into Eqs.~\eqref{prod3} one arrives at 
\begin{equation}
A_s=\frac{1}{3}\sum_{i+j+k=s}A_iA_jA_k+2\delta_{s,1}
\end{equation}
The generating function $\sum_{s\geq 1} A_s z^s$ is a root of cubic polynomial which can be written in the form [cf. Eq.~\eqref{Cz-3}]
\begin{equation}
\label{Az-3}
\sum_{s\geq 1}A_s z^s = 2 \sin\!\left[\frac{\arcsin(3z)}{3}\right]
\end{equation}
This generating function encapsulates the amplitudes $A_s$. We merely mention the asymptotic behavior
\begin{equation}
\label{cs3:asymp}
c_s\simeq \sqrt{\frac{2}{3\pi s^5}}\,\, \frac{3^s}{t^{3s-1}}
\end{equation} 
valid for $t\gg 1$ and odd $s\gg 1$.  

For more comprehensive analysis of the infinite system of differential equations \eqref{prod3}, one can rely again on the recurrent nature of Eqs.~\eqref{prod3}. Using the exponential generating function \eqref{Cz:exp} one reduces an infinite system of equations \eqref{prod3} to a single equation 
\begin{equation}
\label{Burgers:3}
\partial_t \mathcal{E}=\frac{\mathcal{E}^2-t^2}{2}\,\partial_z \mathcal{E}+e^z
\end{equation}
for the generating function. The method of characteristic allows one to recast a hyperbolic partial differential equation \eqref{Burgers:3} to ordinary differential equations, but so far we haven't found even a parametric solution. 

Here we merely show how to determine the moments \eqref{Mn:finite}. Specializing \eqref{Burgers:3} to $z=0$ and using
\begin{equation}
\mathcal{E}\big|_{z=0}=t-g, \qquad \partial_z \mathcal{E}\big|_{z=0}=M_2
\end{equation}
we obtain 
\begin{equation}
\label{M2-3}
M_2 = \frac{2}{2tg-g^2}\,\frac{dg}{dt}
\end{equation}
in the percolating phase; in the pre-percolating phase, the second moment is given by \eqref{M2:sol3}. 
Similarly one expresses the third moment 
\begin{equation}
\label{M3-3}
M_3 = \frac{2}{2tg-g^2}\left[1+(t-g)M_2^2-\frac{dM_2}{dt}\right]
\end{equation}
in the percolating phase via $g$. In the pre-percolating phase, the third moment is given by \eqref{M3:sol3}. 

\subsection{Stockmayer approach}
\label{subsec:SA-3}

The evolution equations 
\begin{equation}
\label{prod3-S}
\frac{d c_s}{dt}=\frac{1}{6}\sum_{i+j+k=s}ijkc_ic_jc_k  - \frac{1}{2}\,M_1^2 sc_s+\delta_{s,1}
\end{equation}
coincide with \eqref{prod} in the pre-percolating phase where $M_1=t$. For $t>t_g$, the mass concentration $M_1(t)$ of finite clusters is a priori unknown. We leave the analysis of relaxation to the future and focus on the steady state, which we probe using the same method as in the binary case (Sec.~\ref{subsec:SA}). Writing again $A_s = sc_s$, we arrive at the recurrence 
\begin{equation}
\label{As-mu3}
\mu^2 A_s=\frac{1}{3}\sum_{i+j+k=s}A_iA_jA_k+2\delta_{s,1}
\end{equation}
in the steady state ($t=\infty$). Using the generating function \eqref{Az} we recast Eqs.~\eqref{As-mu3} into $\mathcal{A}^3-3\mu^2 \mathcal{A}+6z=0$ from which 
\begin{equation*}
\sum_{s\geq 1}A_s z^s = 2\mu \sin\!\left[\frac{\arcsin(3z/\mu^3)}{3}\right]
\end{equation*}
Similarly to the binary case one finds different behaviors depending on whether $\mu$ smaller, equal, or larger than $\sqrt[3]{3}$. The consistent results emerge when
\begin{equation}
\label{M1-3-S}
\mu=M_1(\infty)=\sqrt[3]{3}
\end{equation}
Thus 
\begin{equation}
\label{Az-3-S}
\sum_{s\geq 1}s c_s z^s = 2 \sqrt[3]{3} \sin\!\left[\frac{\arcsin z}{3}\right]
\end{equation}
The concentrations $c_s$ with even $s$ equal to zero. The first four non-vanishing stationary concentrations are
\begin{equation}
\label{1357-S}
\begin{tabular}{cccc}
$c_1=\frac{2\sqrt[3]{3}}{3}$, & $c_3= \frac{8\sqrt[3]{3}}{243}$, & $c_5= \frac{32\sqrt[3]{3}}{3645}$, & $c_7= \frac{512\sqrt[3]{3}}{137781}$
\end{tabular}
\end{equation}
From the behavior of the right-hand side of \eqref{Az-3-S} around $z=1$ we find that for odd $s\gg 1$, the asymptotic is 
\begin{equation}
\label{cs-3-S}
 c_s \simeq 3^{-\frac{1}{6}} \sqrt{\frac{2}{\pi}}\, s^{-5/2}
\end{equation}
To determine the concentration $c=\sum_{s\geq 1}c_s$, we divide \eqref{Az-3-S} by $z$ and integrate over $0<z<1$ to yield
\begin{equation}
\label{c-3-S}
c = \sqrt[3]{3}\,\,\frac{24+\sqrt{3}\ln\big(18817-10864\sqrt{3}\big)}{8}
\end{equation}
Our numerical integration gives value close to the above theoretical prediction $c=1.03692265\ldots$.

\section{Numerical simulations}
\label{sec:numerical}

\subsection{Overview of an algorithm}

We truncate an infinite system to a finite number of differential equations. Such an approximation is applicable as long as mass is conserved (on the numerical level of accuracy). One may then apply any time-integration method for solving the target finite system. As soon as mass conservation law breaks down numerically, we extend the system, i.e., use more equations, and continue the computations. The maximal allowed number of equations cannot exceed a threshold reflecting memory and complexity limitations. Such techniques are often sufficient for the investigation of the asymptotic properties of the solution.

We employ an effective numerical scheme for Smoluchowski equations accounting for binary and ternary aggregation. We rely on an approach developed in several papers. The main idea is based on low-rank approximation of binary and ternary operators with canonical polyadic \cite{bro1997parafac} or tensor train \cite{oseledets2011tensor, oseledets2010tt} decomposition and consecutive evaluation of discrete convolution with fast Fourier transform (FFT) \cite{cooley1965algorithm}. Discrete convolution is a set of sums with the following structure:
\begin{equation}
   c_s=\sum_{i+j = s} a_i b_j,  \quad s = 1,\ldots, N
\end{equation}
where $a$ and $b$ are vectors with $N$ elements. The FFT-based algorithm of convolution computation proceeds by (i) computation of FFT of $a$ and $b$, (ii)  element-wise multiplication, (iii) inverse FFT of the resulting array.

\begin{figure}[ht!]
\centering
\includegraphics[scale=0.6]{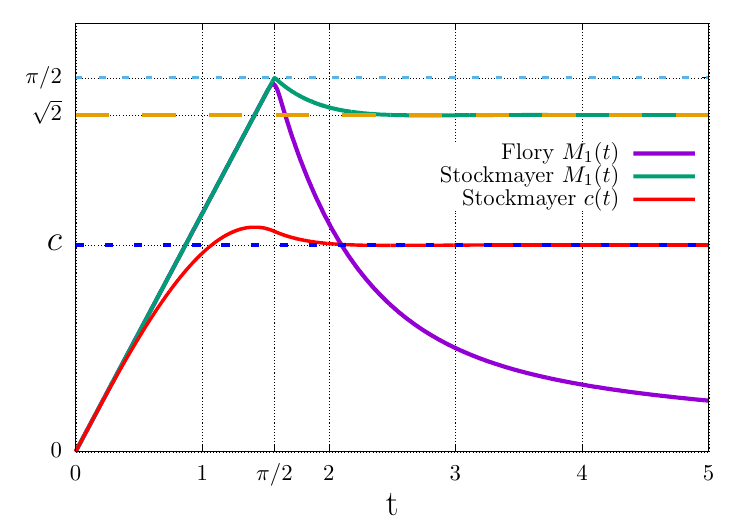}
    \caption{Comparison of $M_1(t)$ for Flory and Stockmayer approaches to input-driven binary aggregation with product kernel.
    In the pre-percolating phase, $M_1=t$ in both cases. In the percolating phase, $M_1$ asymptotically vanishes in the Flory framework,
    namely $M_1\simeq t^{-1}$ in the long time limit. In the Stockmayer framework, $M_1$ approaches to $\sqrt{2}$, and the total cluster density
    also saturates, $c(t)\to c$, with numerical value very close to the theoretical prediction $c=(1-\ln 2)\sqrt{8}$. }
\label{fig:Mass_diff}
\end{figure}

Evaluation of the discrete convolutions with FFT requires $O(N \log N)$ operations \cite{cooley1965algorithm}. The remaining operations on the right-hand side are multiplications of the rank-1 matrices by vectors taking $O(N)$ operations. Such organization of computational process allows one to decrease the formal complexity of evaluation of the right-hand side from $O(N^3)$ operations to $O(N \log N)$. In a previous work \cite{stefonishin2019tensors}, we investigated binary and ternary kernels, $K_{i,j}$ and $K_{i,j,k}$, allowing one to obtain good representations in tensor-train format (both constant and product kernels belong to this class). 

As soon as the complexity of evaluation of the right-hand side is reduced we apply the explicit second-order Runge-Kutta method for solving the system of differential equations \eqref{ckt-23} numerically with modest computational cost for each time-step. 

\subsection{Binary case}

For numerical verification of the stationary solution for the model with mass-independent rates, we used $2^{20}$ equations. We obtained an excellent agreement between the numerical integration of a finite but very large number of equations and an analytical solution, Eq.~\eqref{ck-asymp}, for the infinite system (Fig.~\ref{fig:pure_bin}).

\begin{figure}
 \centering
\includegraphics[scale=0.6]{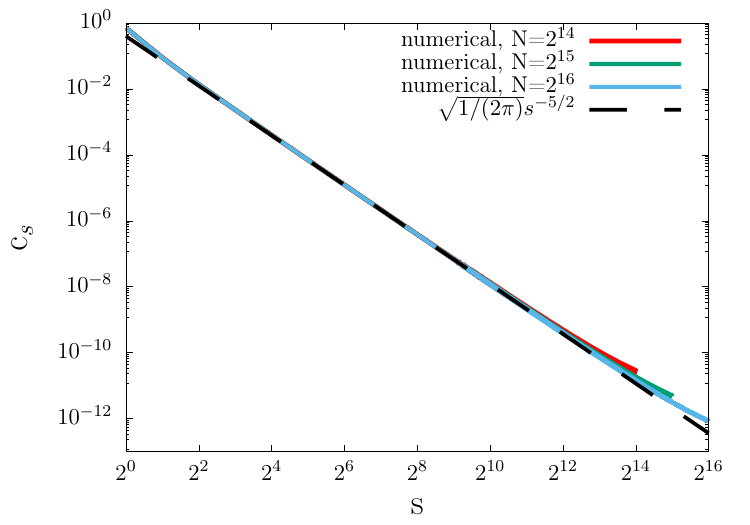}
    \caption{An algebraic decay of concentrations following from numerical integration of Eqs.~\eqref{prod_bad} at $t=5$ with number of 
    equations $N=2^{14}\,, 2^{15}\,, 2^{16}$. The large $s$ asymptotic of the stationary solution, $(2\pi)^{-1/2}s^{-5/2}$, following from the exact
    solution \eqref{cs-S}, is also shown for comparison. }
\label{fig:Strange_cs}
\end{figure}

The numerical investigation of the binary aggregation with product kernel is more subtle than the model with mass-independent rates, or other models where gels do not arise. A natural truncated version of Eqs.~\eqref{Smol} with product kernel is 
\begin{equation}
\label{prod_bad}
\frac{d c_s}{dt}=\frac{1}{2}\sum_{i+j=s}ijc_ic_j-sc_s \sum_{j=1}^{N} j c_j+\delta_{s,1}.
\end{equation}
In Figs.~\ref{fig:Mass_diff}--\ref{fig:Strange_cs} we demonstrate that in the percolating phase ($t>t_g$), numerical integration of Eqs.~\eqref{prod_bad} gives approximations of the solution corresponding to the Stockmayer approach \eqref{prod-S}. Quality of the approximation grows with increase of the size of the truncated system $N$. In our experiments we vary it from $2^{14}$ to $2^{16}$ equations.

In earlier studies of gelation, two ways of describing the evolution in the post-gel regime were suggested by Flory \cite{Flory41} and Stockmayer \cite{Stockmayer43}. The Flory approach accounts for the merging of finite clusters and gel. The Stockmayer approach disregards this phenomenon and it is mute about the gel. Both the Flory and Stockmayer approaches predict the phase transition at the same $t_g$. 

Numerical integrations of truncated systems converge to the predictions of the Flory and Stockmayer approaches for the infinite system. One just ought to truncate Eqs.~\eqref{prod} for the former and Eqs.~\eqref{prod_bad} for the latter. The numerical integration of Eqs.~\eqref{prod_bad} shows the emergence of the steady state. Results of simulations for the mass and cluster concentrations are very close to our analytical predictions \eqref{M1-S} and \eqref{c-S}, see Figs.~\ref{fig:Mass_diff}--\ref{fig:Strange_cs}. The analytical prediction for the stationary mass distribution, Eq.~\eqref{cs-S}, is also in excellent agreement with simulations.

\subsection{Ternary case}

For the truncated version of Eqs.~\eqref{ckt-23} describing aggregation with binary and ternary mass-independent merging events, the convergence of concentrations of the small aggregates to stationary values occurs on a relatively short time and it can be verified with modest computing resources (see Fig.~\ref{fig:C123_dynamics}) using straightforward computations with just three hundred equations. 

Finding the asymptotic behavior of the concentrations is more challenging. In Fig.~\ref{fig:Validation_bin_tern}, we present the results of numerical integration of Eqs.~\eqref{ckt-23} with ternary merging events using $2^{16}$ equations. We obtain the  same asymptotic structure of the solution even though concentrations of the light clusters exhibit significant oscillations when $\lambda \ll 1$. The binary aggregation process dominates for the large aggregates ($s \gg 1$) where oscillations disappear, the solution is monotonically decaying in agreement with theory, and the asymptotic $c_s \sim s^{-3/2}$ becomes manifest.

In Fig.~\ref{fig:general_dynamics}, we show a gradual change of the concentrations for Eqs.~\eqref{ckt-23} coming at the analytical steady state with increasing time from $t=1$ to $t=100$ using $2^{18}$ equations. The apparent decline on the right side of the distributions for both  Fig.~\ref{fig:Validation_bin_tern}--\ref{fig:general_dynamics}  for the large particles stems from a finite number of equations and goes in agreement with crossover rule \eqref{cross}. Stationary concentrations of particles stabilize close to the observed asymptotic rule with further exponential decrease for the masses larger than the crossover rule (see Fig.~\ref{fig:general_dynamics}).

\begin{figure}[ht!]
    \centering
    \includegraphics[scale=0.6]{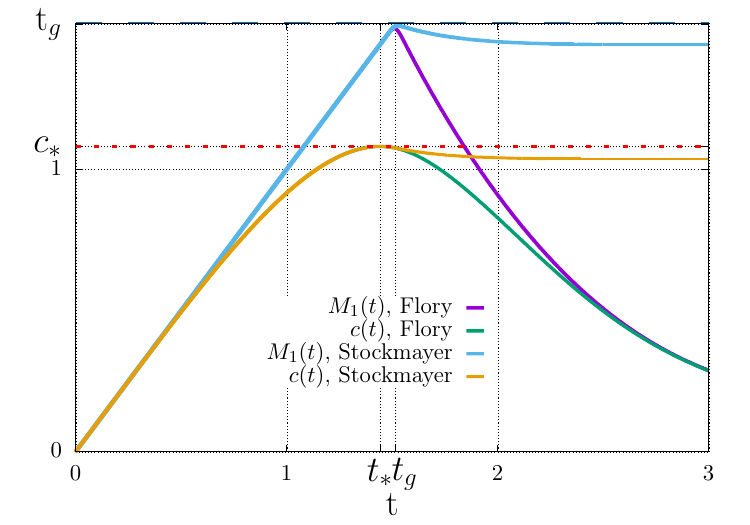}
    \caption{Comparison of $M_1(t)$ and $c(t)$ for the Flory and Stockmayer approaches to input-driven ternary aggregation with product kernel.} 
    \label{fig:ternary_gelation}
\end{figure}

The numerical integration of the ternary aggregation process with product rates is again more subtle. We use a truncated version of Eqs.~\eqref{prod3} in the Flory framework, and a truncated version of Eqs.~\eqref{prod3-S}, that is,
 \begin{equation}
\label{prod3-S-T}
\frac{d c_s}{dt}=\frac{1}{6}\sum_{i+j+k=s}ijkc_ic_jc_k  - \frac{sc_s}{2} \left(\sum_{i\leq N} i c_i\right)^2 +\delta_{s,1}
\end{equation}
with $s=1,\ldots,N$ in the Stockmayer framework. Both truncated versions well approximate an infinite system and give almost identical results (when $N\gg 1$) up to the gelation point (see Fig.~\ref{fig:ternary_gelation}). In the long time limit, however, the numerical integration of Eqs.~\eqref{prod3-S-T} gives concentrations quickly approaching the steady state. The asymptotic behavior in the bulk, namely when $s\gg 1$ and $N-s\gg 1$, is $c_s \sim s^{-5/2}$ for the odd sizes $s$.  The analytical predictions for the steady state, Eqs.~\eqref{M1-3-S}, \eqref{cs-3-S}, and \eqref{c-3-S} are in excellent agreement with simulations.

\section{Discussion}
\label{sec:concl}

We studied aggregation processes driven by the source of monomers. For the aggregation process with mass-independent merging rates, the well-known $c_s \sim s^{-3/2}$ asymptotic can be generalized from the binary aggregation to the model involving ternary merging events. One can similarly generalize to higher-order merging events. The exact and asymptotic expressions for the stationary concentrations are useful for the verification of miscellaneous numerical methods. For the model with mass-independent merging rates, theory and simulations are in excellent agreement.

The input-driven aggregation with product aggregation rates is particularly interesting and challenging because of the phase transition known as gelation. The exact gelation time $t_g$ for the initially empty system is given by Eq.~\eqref{tg:2} for the binary aggregation and by Eq.~\eqref{tg:3} for pure ternary aggregation. For pure $n-$ary aggregation, starting with
\begin{equation*}
\frac{d c_s}{dt}=\sum_{i_1+\ldots+i_n=s}\frac{i_1\ldots i_nc_{i_1}\ldots c_{i_n}}{n(n-1)}  - \frac{t^{n-1}}{n-1}\,sc_s +\delta_{s,1}
\end{equation*}
one deduces the equation for the second moment
\begin{equation}
\label{M2:n}
\frac{d M_2}{dt}=1+t^{n-2}M_2^2
\end{equation}
in the pre-percolation phase. The solution of \eqref{M2:n} diverges: $M_2\simeq t_g^{2-n} (t_g-t)^{-1}$ as $t_g-t\to +0$. We know that Eq.~\eqref{M2:n} admits an analytical solution in the binary case, Eq.~\eqref{M2:sol}, and ternary case, Eq.~\eqref{M2:sol3}. We haven't succeeded in expressing the solution of  Eq.~\eqref{M2:n} subject to $M_2(0)=0$ via standard special functions when $n\geq 4$. One can still determine $t_g$ by solving Eq.~\eqref{M2:n} numerically, or approximately using an exact series expansion
\begin{eqnarray*}
M_2 &=& t + \frac{t^{1+n}}{1+n}+\frac{2t^{1+2n}}{(1+n)(1+2n)}\\
&+&\frac{(5+6n)t^{1+3n}}{(1+n)^2(1+2n)(1+3n)}\\
&+& \frac{2(7+12n)t^{1+4n}}{(1+n)^2(1+2n)(1+3n)(1+4n)}\\
&+& \frac{2(21+118n+214n^2+120 n^3)t^{1+5n}}{(1+n)^3(1+2n)^2(1+3n)(1+4n)(1+5n)}+\ldots
\end{eqnarray*}
This expansion simplifies to $t(1-n^{-1}t^n)^{-1}$ for large $n$, from which $t_g\to n^{1/n}$, or $t_g-1\to n^{-1}\ln n$ when $n\to\infty$. 

We analyzed the input-driven binary and ternary aggregation processes with the product kernel, and we also analyzed the problem numerically. One ought to distinguish two possible ways of evolution in the post-percolation phase. A steady state is formed in the realm of the Stockmayer approach. The analytical predictions for the steady state, Eqs.~\eqref{M1-S}--\eqref{c-S} in the binary case and Eqs.~\eqref{M1-3-S}--\eqref{c-3-S} in the ternary case, are well captured by numerical integration.

An interesting challenge is to study other aggregation processes exhibiting gelation. We only mention a one-parameter family of models with generalized product kernel $K_{i,j}= (ij)^{\nu}$. Gelation occurs when  $\frac{1}{2}<\nu \leq 1$. (Instantaneous gelation occurs when $\nu>1$.) The governing equations read 
\begin{equation}
\label{prod-nu}
\frac{d c_s}{dt}=\frac{1}{2}\sum_{i+j=s} (ij)^{\nu} c_ic_j-s^\nu c_s M_\nu +\delta_{s,1}
\end{equation}
where $M_\nu = \sum_{j\geq 1} j^\nu c_j$ is the  $\nu^\text{th}$ moment. The product kernel ($\nu=1$) is very special. We know the gelation time only in this case. The exact value of $M_\nu(t)$ which is necessary for the analysis of the Flory version is also known only when $\nu=1$. Therefore, it is unclear how to write an appropriate truncated version of \eqref{prod-nu} in the Flory framework. In contrast, the steady-state emerging in terms of Stockmayer is easy to extract for arbitrary $\nu$. One arrives at the same recurrence \eqref{As-mu:rec} as before, and the only difference is that now $A_s = s^\nu c_s$ and $\mu=M_\nu(\infty)$. Therefore
\begin{subequations}
\begin{align}
\label{M-nu-inf}
&\mu = M_\nu(\infty) = \sqrt{2}\\
\label{cs-nu-S}
&c_s=\frac{1}{\sqrt{2\pi}}\,\,
\frac{\Gamma(s-\frac{1}{2})}{s^\nu\Gamma(s+1)}
\end{align}
\end{subequations}

The generalized product kernel and a few other kernels have been studied numerically using a less accurate coarse-graining approach \cite{ball2011instantaneous, lee2000validity}. Extending our numerical treatment to such kernels using the Flory framework is left for the future. 

We relied on the Smoluchowski equations providing the mean-field description of the well-mixed system. Extending our analyses to diffusion-controlled aggregation processes in low spatial dimensions where the mean-field description fails \cite{book} is the major challenge requiring different analytical and numerical tools. Even well-mixed systems undergo fluctuations when the total mass is finite. Numerical integration methods are inadequate for probing fluctuations and large deviations. Monte Carlo methods might shed light on rare events \cite{dandekar2023monte} or finite size effects \cite{dyachenko2023finite}.

\bigskip\noindent
S. M. was partly supported by Moscow Center for Fundamental and Applied Mathematics at INM RAS, Ministry of Education and Science of the Russian Federation (Grant No. 075-15-2022-286). S. M. would like to thank Nikolai Brilliantov and 	Alexander Osinsky for useful discussions in the beginning of this study.

\bibliography{TripleSmol}

\begin{thebibliography}{79}%
\makeatletter
\providecommand \@ifxundefined [1]{%
 \@ifx{#1\undefined}
}%
\providecommand \@ifnum [1]{%
 \ifnum #1\expandafter \@firstoftwo
 \else \expandafter \@secondoftwo
 \fi
}%
\providecommand \@ifx [1]{%
 \ifx #1\expandafter \@firstoftwo
 \else \expandafter \@secondoftwo
 \fi
}%
\providecommand \natexlab [1]{#1}%
\providecommand \enquote  [1]{``#1''}%
\providecommand \bibnamefont  [1]{#1}%
\providecommand \bibfnamefont [1]{#1}%
\providecommand \citenamefont [1]{#1}%
\providecommand \href@noop [0]{\@secondoftwo}%
\providecommand \href [0]{\begingroup \@sanitize@url \@href}%
\providecommand \@href[1]{\@@startlink{#1}\@@href}%
\providecommand \@@href[1]{\endgroup#1\@@endlink}%
\providecommand \@sanitize@url [0]{\catcode `\\12\catcode `\$12\catcode
  `\&12\catcode `\#12\catcode `\^12\catcode `\_12\catcode `\%12\relax}%
\providecommand \@@startlink[1]{}%
\providecommand \@@endlink[0]{}%
\providecommand \url  [0]{\begingroup\@sanitize@url \@url }%
\providecommand \@url [1]{\endgroup\@href {#1}{\urlprefix }}%
\providecommand \urlprefix  [0]{URL }%
\providecommand \Eprint [0]{\href }%
\providecommand \doibase [0]{https://doi.org/}%
\providecommand \selectlanguage [0]{\@gobble}%
\providecommand \bibinfo  [0]{\@secondoftwo}%
\providecommand \bibfield  [0]{\@secondoftwo}%
\providecommand \translation [1]{[#1]}%
\providecommand \BibitemOpen [0]{}%
\providecommand \bibitemStop [0]{}%
\providecommand \bibitemNoStop [0]{.\EOS\space}%
\providecommand \EOS [0]{\spacefactor3000\relax}%
\providecommand \BibitemShut  [1]{\csname bibitem#1\endcsname}%
\let\auto@bib@innerbib\@empty
\bibitem [{\citenamefont {Flory}(1953)}]{Flory53}%
  \BibitemOpen
  \bibfield  {author} {\bibinfo {author} {\bibfnamefont {P.~J.}\ \bibnamefont
  {Flory}},\ }\href@noop {} {\emph {\bibinfo {title} {Principles of {P}olymer
  {C}hemistry}}}\ (\bibinfo  {publisher} {Cornell University Press},\ \bibinfo
  {address} {New York},\ \bibinfo {year} {1953})\BibitemShut {NoStop}%
\bibitem [{\citenamefont {Krapivsky}\ \emph {et~al.}(2010)\citenamefont
  {Krapivsky}, \citenamefont {Redner},\ and\ \citenamefont {Ben-Naim}}]{book}%
  \BibitemOpen
  \bibfield  {author} {\bibinfo {author} {\bibfnamefont {P.~L.}\ \bibnamefont
  {Krapivsky}}, \bibinfo {author} {\bibfnamefont {S.}~\bibnamefont {Redner}},\
  and\ \bibinfo {author} {\bibfnamefont {E.}~\bibnamefont {Ben-Naim}},\ }\href
  {https://doi.org/10.1017/CBO9780511780516} {\emph {\bibinfo {title} {A
  kinetic view of statistical physics}}}\ (\bibinfo  {publisher} {Cambridge
  University Press},\ \bibinfo {year} {2010})\BibitemShut {NoStop}%
\bibitem [{\citenamefont {Leyvraz}(2003)}]{Leyvraz03}%
  \BibitemOpen
  \bibfield  {author} {\bibinfo {author} {\bibfnamefont {F.}~\bibnamefont
  {Leyvraz}},\ }\bibfield  {title} {\bibinfo {title} {Scaling theory and
  exactly solved models in the kinetics of irreversible aggregation},\ }\href
  {https://doi.org/10.1016/S0370-1573(03)00241-2} {\bibfield  {journal}
  {\bibinfo  {journal} {Physics Reports}\ }\textbf {\bibinfo {volume} {383}},\
  \bibinfo {pages} {95} (\bibinfo {year} {2003})}\BibitemShut {NoStop}%
\bibitem [{\citenamefont {Anand}\ \emph {et~al.}(2006)\citenamefont {Anand},
  \citenamefont {Rajagopal},\ and\ \citenamefont {Rajagopal}}]{anand2006model}%
  \BibitemOpen
  \bibfield  {author} {\bibinfo {author} {\bibfnamefont {M.}~\bibnamefont
  {Anand}}, \bibinfo {author} {\bibfnamefont {K.}~\bibnamefont {Rajagopal}},\
  and\ \bibinfo {author} {\bibfnamefont {K.~R.}\ \bibnamefont {Rajagopal}},\
  }\bibfield  {title} {\bibinfo {title} {A model for the formation and lysis of
  blood clots},\ }\href {https://doi.org/10.1159/000089931} {\bibfield
  {journal} {\bibinfo  {journal} {Pathophysiology of haemostasis and
  thrombosis}\ }\textbf {\bibinfo {volume} {34}},\ \bibinfo {pages} {109}
  (\bibinfo {year} {2006})}\BibitemShut {NoStop}%
\bibitem [{\citenamefont {Voloshchuk}\ and\ \citenamefont
  {Sedunov}(1977)}]{voloshchuk1977kinetics}%
  \BibitemOpen
  \bibfield  {author} {\bibinfo {author} {\bibfnamefont {V.~M.}\ \bibnamefont
  {Voloshchuk}}\ and\ \bibinfo {author} {\bibfnamefont {Y.~S.}\ \bibnamefont
  {Sedunov}},\ }\bibfield  {title} {\bibinfo {title} {Kinetics of
  condensation-coagulation processes in the {E}arth atmosphere},\ }\href
  {http://mi.mathnet.ru/dan41083} {\bibfield  {journal} {\bibinfo  {journal}
  {Dokl. Akad. Nauk}\ }\textbf {\bibinfo {volume} {235}},\ \bibinfo {pages}
  {50} (\bibinfo {year} {1977})}\BibitemShut {NoStop}%
\bibitem [{\citenamefont {Chokshi}\ \emph {et~al.}(1993)\citenamefont
  {Chokshi}, \citenamefont {Tielens},\ and\ \citenamefont
  {Hollenbach}}]{dust93}%
  \BibitemOpen
  \bibfield  {author} {\bibinfo {author} {\bibfnamefont {A.}~\bibnamefont
  {Chokshi}}, \bibinfo {author} {\bibfnamefont {A.~G.~G.}\ \bibnamefont
  {Tielens}},\ and\ \bibinfo {author} {\bibfnamefont {D.}~\bibnamefont
  {Hollenbach}},\ }\bibfield  {title} {\bibinfo {title} {Dust coagulation},\
  }\href {https://doi.org/10.1086/172562} {\bibfield  {journal} {\bibinfo
  {journal} {Astrophys. J.}\ }\textbf {\bibinfo {volume} {407}},\ \bibinfo
  {pages} {806} (\bibinfo {year} {1993})}\BibitemShut {NoStop}%
\bibitem [{\citenamefont {Aloyan}\ \emph {et~al.}(1997)\citenamefont {Aloyan},
  \citenamefont {Arutyunyan}, \citenamefont {Lushnikov},\ and\ \citenamefont
  {Zagaynov}}]{aloyan1997transport}%
  \BibitemOpen
  \bibfield  {author} {\bibinfo {author} {\bibfnamefont {A.~E.}\ \bibnamefont
  {Aloyan}}, \bibinfo {author} {\bibfnamefont {V.~O.}\ \bibnamefont
  {Arutyunyan}}, \bibinfo {author} {\bibfnamefont {A.~A.}\ \bibnamefont
  {Lushnikov}},\ and\ \bibinfo {author} {\bibfnamefont {V.~A.}\ \bibnamefont
  {Zagaynov}},\ }\bibfield  {title} {\bibinfo {title} {Transport of coagulating
  aerosol in the atmosphere},\ }\href
  {https://doi.org/10.1016/S0021-8502(96)00043-2} {\bibfield  {journal}
  {\bibinfo  {journal} {Journal of aerosol science}\ }\textbf {\bibinfo
  {volume} {28}},\ \bibinfo {pages} {67} (\bibinfo {year} {1997})}\BibitemShut
  {NoStop}%
\bibitem [{\citenamefont {Friedlander}(2000)}]{Friedlander}%
  \BibitemOpen
  \bibfield  {author} {\bibinfo {author} {\bibfnamefont {S.~K.}\ \bibnamefont
  {Friedlander}},\ }\href@noop {} {\emph {\bibinfo {title} {Smoke, Dust and
  Haze}}}\ (\bibinfo  {publisher} {Oxford University Press},\ \bibinfo
  {address} {Oxford},\ \bibinfo {year} {2000})\BibitemShut {NoStop}%
\bibitem [{\citenamefont {Lushnikov}(2010)}]{lushnikov2010introduction}%
  \BibitemOpen
  \bibfield  {author} {\bibinfo {author} {\bibfnamefont {A.~A.}\ \bibnamefont
  {Lushnikov}},\ }\href {https://doi.org/10.1002/9783527630134} {\emph
  {\bibinfo {title} {Introduction to aerosols}}}\ (\bibinfo  {publisher} {Wiley
  Online Library},\ \bibinfo {year} {2010})\ pp.\ \bibinfo {pages}
  {1--41}\BibitemShut {NoStop}%
\bibitem [{\citenamefont {Kaganer}\ \emph {et~al.}(2016)\citenamefont
  {Kaganer}, \citenamefont {Fernandez-Garrido}, \citenamefont {Dogan},
  \citenamefont {Sabelfeld},\ and\ \citenamefont
  {Brandt}}]{kaganer2016nucleation}%
  \BibitemOpen
  \bibfield  {author} {\bibinfo {author} {\bibfnamefont {V.~M.}\ \bibnamefont
  {Kaganer}}, \bibinfo {author} {\bibfnamefont {S.}~\bibnamefont
  {Fernandez-Garrido}}, \bibinfo {author} {\bibfnamefont {P.}~\bibnamefont
  {Dogan}}, \bibinfo {author} {\bibfnamefont {K.~K.}\ \bibnamefont
  {Sabelfeld}},\ and\ \bibinfo {author} {\bibfnamefont {O.}~\bibnamefont
  {Brandt}},\ }\bibfield  {title} {\bibinfo {title} {Nucleation, growth, and
  bundling of {G}a{N} nanowires in molecular beam epitaxy: {D}isentangling the
  origin of nanowire coalescence},\ }\href
  {https://doi.org/10.1021/acs.nanolett.6b01044} {\bibfield  {journal}
  {\bibinfo  {journal} {Nano Lett.}\ }\textbf {\bibinfo {volume} {16}},\
  \bibinfo {pages} {3717} (\bibinfo {year} {2016})}\BibitemShut {NoStop}%
\bibitem [{\citenamefont {Hinnemann}\ \emph {et~al.}(2003)\citenamefont
  {Hinnemann}, \citenamefont {Hinrichsen},\ and\ \citenamefont
  {Wolf}}]{hinnemann2003epitaxial}%
  \BibitemOpen
  \bibfield  {author} {\bibinfo {author} {\bibfnamefont {B.}~\bibnamefont
  {Hinnemann}}, \bibinfo {author} {\bibfnamefont {H.}~\bibnamefont
  {Hinrichsen}},\ and\ \bibinfo {author} {\bibfnamefont {D.~E.}\ \bibnamefont
  {Wolf}},\ }\bibfield  {title} {\bibinfo {title} {Epitaxial growth with pulsed
  deposition: {S}ubmonolayer scaling and {V}illain instability},\ }\href
  {https://doi.org/10.1103/PhysRevE.67.011602} {\bibfield  {journal} {\bibinfo
  {journal} {Phys. Rev. E}\ }\textbf {\bibinfo {volume} {67}},\ \bibinfo
  {pages} {011602} (\bibinfo {year} {2003})}\BibitemShut {NoStop}%
\bibitem [{\citenamefont {Esposito}(2006)}]{Esposito}%
  \BibitemOpen
  \bibfield  {author} {\bibinfo {author} {\bibfnamefont {L.}~\bibnamefont
  {Esposito}},\ }\href@noop {} {\emph {\bibinfo {title} {Planetary Rings}}}\
  (\bibinfo  {publisher} {Cambridge University Press},\ \bibinfo {address}
  {Cambridge, UK},\ \bibinfo {year} {2006})\BibitemShut {NoStop}%
\bibitem [{\citenamefont {G\"{u}ttler}\ \emph {et~al.}(2010)\citenamefont
  {G\"{u}ttler}, \citenamefont {Blum}, \citenamefont {Zsom}, \citenamefont
  {Ormel},\ and\ \citenamefont {Dullemond}}]{Guettler10}%
  \BibitemOpen
  \bibfield  {author} {\bibinfo {author} {\bibfnamefont {C.}~\bibnamefont
  {G\"{u}ttler}}, \bibinfo {author} {\bibfnamefont {J.}~\bibnamefont {Blum}},
  \bibinfo {author} {\bibfnamefont {A.}~\bibnamefont {Zsom}}, \bibinfo {author}
  {\bibfnamefont {C.}~\bibnamefont {Ormel}},\ and\ \bibinfo {author}
  {\bibfnamefont {C.~P.}\ \bibnamefont {Dullemond}},\ }\bibfield  {title}
  {\bibinfo {title} {The outcome of protoplanetary dust growth: {P}ebbles,
  boulders, or planetesimals?},\ }\href
  {https://doi.org/10.1051/0004-6361/200912852} {\bibfield  {journal} {\bibinfo
   {journal} {A \& A}\ }\textbf {\bibinfo {volume} {513}},\ \bibinfo {pages}
  {A56} (\bibinfo {year} {2010})}\BibitemShut {NoStop}%
\bibitem [{\citenamefont {Schr\"{a}pler}\ and\ \citenamefont
  {Blum}(2011)}]{Blum11}%
  \BibitemOpen
  \bibfield  {author} {\bibinfo {author} {\bibfnamefont {R.}~\bibnamefont
  {Schr\"{a}pler}}\ and\ \bibinfo {author} {\bibfnamefont {J.}~\bibnamefont
  {Blum}},\ }\bibfield  {title} {\bibinfo {title} {The physics of
  protopanetesimal dust agglomerates. {VI}. {E}rosion of large aggregates as a
  source of micrometer-sized particles},\ }\href
  {https://dx.doi.org/10.1088/0004-637X/734/2/108} {\bibfield  {journal}
  {\bibinfo  {journal} {Astrophys. J.}\ }\textbf {\bibinfo {volume} {734}},\
  \bibinfo {pages} {108} (\bibinfo {year} {2011})}\BibitemShut {NoStop}%
\bibitem [{\citenamefont {Brilliantov}\ \emph {et~al.}(2015)\citenamefont
  {Brilliantov}, \citenamefont {Krapivsky}, \citenamefont {Bodrova},
  \citenamefont {Spahn}, \citenamefont {Hayakawa}, \citenamefont {Stadnichuk},\
  and\ \citenamefont {Schmidt}}]{brilliantov2015size}%
  \BibitemOpen
  \bibfield  {author} {\bibinfo {author} {\bibfnamefont {N.}~\bibnamefont
  {Brilliantov}}, \bibinfo {author} {\bibfnamefont {P.~L.}\ \bibnamefont
  {Krapivsky}}, \bibinfo {author} {\bibfnamefont {A.}~\bibnamefont {Bodrova}},
  \bibinfo {author} {\bibfnamefont {F.}~\bibnamefont {Spahn}}, \bibinfo
  {author} {\bibfnamefont {H.}~\bibnamefont {Hayakawa}}, \bibinfo {author}
  {\bibfnamefont {V.}~\bibnamefont {Stadnichuk}},\ and\ \bibinfo {author}
  {\bibfnamefont {J.}~\bibnamefont {Schmidt}},\ }\bibfield  {title} {\bibinfo
  {title} {Size distribution of particles in {S}aturn's rings from aggregation
  and fragmentation},\ }\href {https://doi.org/10.1073/pnas.1503957112}
  {\bibfield  {journal} {\bibinfo  {journal} {PNAS}\ }\textbf {\bibinfo
  {volume} {112}},\ \bibinfo {pages} {9536} (\bibinfo {year}
  {2015})}\BibitemShut {NoStop}%
\bibitem [{\citenamefont {Smoluchowski}(1916)}]{smoluchowski1916drei}%
  \BibitemOpen
  \bibfield  {author} {\bibinfo {author} {\bibfnamefont {M.~v.}\ \bibnamefont
  {Smoluchowski}},\ }\bibfield  {title} {\bibinfo {title} {Drei vortrage uber
  diffusion, {B}rownsche bewegung und koagulation von kolloidteilchen},\
  }\href@noop {} {\bibfield  {journal} {\bibinfo  {journal} {Zeit. Physik}\
  }\textbf {\bibinfo {volume} {17}},\ \bibinfo {pages} {557} (\bibinfo {year}
  {1916})}\BibitemShut {NoStop}%
\bibitem [{\citenamefont {Lushnikov}(1973)}]{lushnikov1973evolution}%
  \BibitemOpen
  \bibfield  {author} {\bibinfo {author} {\bibfnamefont {A.~A.}\ \bibnamefont
  {Lushnikov}},\ }\bibfield  {title} {\bibinfo {title} {Evolution of
  coagulating systems},\ }\href {https://doi.org/10.1016/0021-9797(73)90171-9}
  {\bibfield  {journal} {\bibinfo  {journal} {Journal of Colloid and Interface
  Science}\ }\textbf {\bibinfo {volume} {45}},\ \bibinfo {pages} {549}
  (\bibinfo {year} {1973})}\BibitemShut {NoStop}%
\bibitem [{\citenamefont {Galkin}(2001)}]{galkin2001}%
  \BibitemOpen
  \bibfield  {author} {\bibinfo {author} {\bibfnamefont {V.}~\bibnamefont
  {Galkin}},\ }\bibfield  {title} {\bibinfo {title} {Smoluchowski equation (in
  russian)},\ }\href {https://www.rfbr.ru/rffi/portal/books/o_18655} {\bibfield
   {journal} {\bibinfo  {journal} {Moscow, RF, Fizmatlit}\ } (\bibinfo {year}
  {2001})}\BibitemShut {NoStop}%
\bibitem [{\citenamefont {Matveev}\ \emph {et~al.}(2017)\citenamefont
  {Matveev}, \citenamefont {Krapivsky}, \citenamefont {Smirnov}, \citenamefont
  {Tyrtyshnikov},\ and\ \citenamefont {Brilliantov}}]{matveev2017oscillations}%
  \BibitemOpen
  \bibfield  {author} {\bibinfo {author} {\bibfnamefont {S.~A.}\ \bibnamefont
  {Matveev}}, \bibinfo {author} {\bibfnamefont {P.~L.}\ \bibnamefont
  {Krapivsky}}, \bibinfo {author} {\bibfnamefont {A.~P.}\ \bibnamefont
  {Smirnov}}, \bibinfo {author} {\bibfnamefont {E.~E.}\ \bibnamefont
  {Tyrtyshnikov}},\ and\ \bibinfo {author} {\bibfnamefont {N.~V.}\ \bibnamefont
  {Brilliantov}},\ }\bibfield  {title} {\bibinfo {title} {Oscillations in
  aggregation-shattering processes},\ }\href
  {https://doi.org/10.1103/PhysRevLett.119.260601} {\bibfield  {journal}
  {\bibinfo  {journal} {Phys. Rev. Lett.}\ }\textbf {\bibinfo {volume} {119}},\
  \bibinfo {pages} {260601} (\bibinfo {year} {2017})}\BibitemShut {NoStop}%
\bibitem [{\citenamefont {Brilliantov}\ \emph {et~al.}(2021)\citenamefont
  {Brilliantov}, \citenamefont {Otieno},\ and\ \citenamefont
  {Krapivsky}}]{Wendy21}%
  \BibitemOpen
  \bibfield  {author} {\bibinfo {author} {\bibfnamefont {N.~V.}\ \bibnamefont
  {Brilliantov}}, \bibinfo {author} {\bibfnamefont {W.}~\bibnamefont
  {Otieno}},\ and\ \bibinfo {author} {\bibfnamefont {P.~L.}\ \bibnamefont
  {Krapivsky}},\ }\bibfield  {title} {\bibinfo {title} {Nonextensive
  supercluster states in aggregation with fragmentation},\ }\href
  {https://doi.org/10.1103/PhysRevLett.127.250602} {\bibfield  {journal}
  {\bibinfo  {journal} {Phys. Rev. Lett.}\ }\textbf {\bibinfo {volume} {127}},\
  \bibinfo {pages} {250602} (\bibinfo {year} {2021})}\BibitemShut {NoStop}%
\bibitem [{\citenamefont {Otieno}\ \emph {et~al.}(2023)\citenamefont {Otieno},
  \citenamefont {Brilliantov},\ and\ \citenamefont {Krapivsky}}]{Wendy23}%
  \BibitemOpen
  \bibfield  {author} {\bibinfo {author} {\bibfnamefont {W.}~\bibnamefont
  {Otieno}}, \bibinfo {author} {\bibfnamefont {N.~V.}\ \bibnamefont
  {Brilliantov}},\ and\ \bibinfo {author} {\bibfnamefont {P.~L.}\ \bibnamefont
  {Krapivsky}},\ }\bibfield  {title} {\bibinfo {title} {Supercluster states and
  phase transitions in aggregation-fragmentation processes},\ }\href
  {https://doi.org/10.1103/PhysRevE.108.044142} {\bibfield  {journal} {\bibinfo
   {journal} {Phys. Rev. E}\ }\textbf {\bibinfo {volume} {108}},\ \bibinfo
  {pages} {044142} (\bibinfo {year} {2023})}\BibitemShut {NoStop}%
\bibitem [{\citenamefont {Bodrova}\ \emph {et~al.}(2019)\citenamefont
  {Bodrova}, \citenamefont {Stadnichuk}, \citenamefont {Krapivsky},
  \citenamefont {Schmidt},\ and\ \citenamefont
  {Brilliantov}}]{bodrova2019kinetic}%
  \BibitemOpen
  \bibfield  {author} {\bibinfo {author} {\bibfnamefont {A.~S.}\ \bibnamefont
  {Bodrova}}, \bibinfo {author} {\bibfnamefont {V.}~\bibnamefont {Stadnichuk}},
  \bibinfo {author} {\bibfnamefont {P.~L.}\ \bibnamefont {Krapivsky}}, \bibinfo
  {author} {\bibfnamefont {J.}~\bibnamefont {Schmidt}},\ and\ \bibinfo {author}
  {\bibfnamefont {N.~V.}\ \bibnamefont {Brilliantov}},\ }\bibfield  {title}
  {\bibinfo {title} {Kinetic regimes in aggregating systems with spontaneous
  and collisional fragmentation},\ }\href
  {https://doi.org/10.1088/1751-8121/ab1616} {\bibfield  {journal} {\bibinfo
  {journal} {J. Phys. A}\ }\textbf {\bibinfo {volume} {52}},\ \bibinfo {pages}
  {205001} (\bibinfo {year} {2019})}\BibitemShut {NoStop}%
\bibitem [{\citenamefont {Ke}\ and\ \citenamefont {Lin}(2002)}]{Ke02}%
  \BibitemOpen
  \bibfield  {author} {\bibinfo {author} {\bibfnamefont {J.}~\bibnamefont
  {Ke}}\ and\ \bibinfo {author} {\bibfnamefont {Z.}~\bibnamefont {Lin}},\
  }\bibfield  {title} {\bibinfo {title} {Kinetics of migration-driven
  aggregation processes},\ }\href {https://doi.org/10.1103/PhysRevE.66.050102}
  {\bibfield  {journal} {\bibinfo  {journal} {Phys. Rev. E}\ }\textbf {\bibinfo
  {volume} {66}},\ \bibinfo {pages} {050102} (\bibinfo {year}
  {2002})}\BibitemShut {NoStop}%
\bibitem [{\citenamefont {Ben-Naim}\ and\ \citenamefont
  {Krapivsky}(2003)}]{PK_exchange03}%
  \BibitemOpen
  \bibfield  {author} {\bibinfo {author} {\bibfnamefont {E.}~\bibnamefont
  {Ben-Naim}}\ and\ \bibinfo {author} {\bibfnamefont {P.~L.}\ \bibnamefont
  {Krapivsky}},\ }\bibfield  {title} {\bibinfo {title} {Exchange-driven
  growth},\ }\href {https://doi.org/10.1103/PhysRevE.68.031104} {\bibfield
  {journal} {\bibinfo  {journal} {Phys. Rev. E}\ }\textbf {\bibinfo {volume}
  {68}},\ \bibinfo {pages} {031104} (\bibinfo {year} {2003})}\BibitemShut
  {NoStop}%
\bibitem [{\citenamefont {Krapivsky}(2015)}]{PK_exchange15}%
  \BibitemOpen
  \bibfield  {author} {\bibinfo {author} {\bibfnamefont {P.~L.}\ \bibnamefont
  {Krapivsky}},\ }\bibfield  {title} {\bibinfo {title} {Mass exchange processes
  with input},\ }\href {https://doi.org/10.1088/1751-8113/48/20/205003}
  {\bibfield  {journal} {\bibinfo  {journal} {J. Phys. A}\ }\textbf {\bibinfo
  {volume} {48}},\ \bibinfo {pages} {205003} (\bibinfo {year}
  {2015})}\BibitemShut {NoStop}%
\bibitem [{\citenamefont {Pego}\ and\ \citenamefont
  {Vel{\'a}zquez}(2020)}]{pego2020temporal}%
  \BibitemOpen
  \bibfield  {author} {\bibinfo {author} {\bibfnamefont {R.~L.}\ \bibnamefont
  {Pego}}\ and\ \bibinfo {author} {\bibfnamefont {J.~J.~L.}\ \bibnamefont
  {Vel{\'a}zquez}},\ }\bibfield  {title} {\bibinfo {title} {Temporal
  oscillations in {B}ecker--{D}{\"o}ring equations with atomization},\ }\href
  {https://doi.org/10.1088/1361-6544/ab6815} {\bibfield  {journal} {\bibinfo
  {journal} {Nonlinearity}\ }\textbf {\bibinfo {volume} {33}},\ \bibinfo
  {pages} {1812} (\bibinfo {year} {2020})}\BibitemShut {NoStop}%
\bibitem [{\citenamefont {Hayakawa}(1987)}]{hayakawa87}%
  \BibitemOpen
  \bibfield  {author} {\bibinfo {author} {\bibfnamefont {H.}~\bibnamefont
  {Hayakawa}},\ }\bibfield  {title} {\bibinfo {title} {Irreversible kinetic
  coagulations in the presence of a source},\ }\href
  {https://doi.org/10.1088/0305-4470/20/12/009} {\bibfield  {journal} {\bibinfo
   {journal} {J. Phys. A}\ }\textbf {\bibinfo {volume} {20}},\ \bibinfo {pages}
  {L801} (\bibinfo {year} {1987})}\BibitemShut {NoStop}%
\bibitem [{\citenamefont {Esenturk}\ and\ \citenamefont
  {Connaughton}(2020)}]{esenturk2020role}%
  \BibitemOpen
  \bibfield  {author} {\bibinfo {author} {\bibfnamefont {E.}~\bibnamefont
  {Esenturk}}\ and\ \bibinfo {author} {\bibfnamefont {C.}~\bibnamefont
  {Connaughton}},\ }\bibfield  {title} {\bibinfo {title} {Role of zero clusters
  in exchange-driven growth with and without input},\ }\href
  {https://doi.org/10.1103/PhysRevE.101.052134} {\bibfield  {journal} {\bibinfo
   {journal} {Phys. Rev. E}\ }\textbf {\bibinfo {volume} {101}},\ \bibinfo
  {pages} {052134} (\bibinfo {year} {2020})}\BibitemShut {NoStop}%
\bibitem [{\citenamefont {Osinsky}\ and\ \citenamefont
  {Brilliantov}(2022)}]{osinsky2022anomalous}%
  \BibitemOpen
  \bibfield  {author} {\bibinfo {author} {\bibfnamefont {A.~I.}\ \bibnamefont
  {Osinsky}}\ and\ \bibinfo {author} {\bibfnamefont {N.~V.}\ \bibnamefont
  {Brilliantov}},\ }\bibfield  {title} {\bibinfo {title} {Anomalous aggregation
  regimes of temperature-dependent {S}moluchowski equations},\ }\href
  {https://doi.org/10.1103/PhysRevE.105.034119} {\bibfield  {journal} {\bibinfo
   {journal} {Phys. Rev. E}\ }\textbf {\bibinfo {volume} {105}},\ \bibinfo
  {pages} {034119} (\bibinfo {year} {2022})}\BibitemShut {NoStop}%
\bibitem [{\citenamefont {Budzinskiy}\ \emph {et~al.}(2021)\citenamefont
  {Budzinskiy}, \citenamefont {Matveev},\ and\ \citenamefont
  {Krapivsky}}]{budzinskiy2021hopf}%
  \BibitemOpen
  \bibfield  {author} {\bibinfo {author} {\bibfnamefont {S.~S.}\ \bibnamefont
  {Budzinskiy}}, \bibinfo {author} {\bibfnamefont {S.~A.}\ \bibnamefont
  {Matveev}},\ and\ \bibinfo {author} {\bibfnamefont {P.~L.}\ \bibnamefont
  {Krapivsky}},\ }\bibfield  {title} {\bibinfo {title} {Hopf bifurcation in
  addition-shattering kinetics},\ }\href
  {https://doi.org/10.1103/PhysRevE.103.L040101} {\bibfield  {journal}
  {\bibinfo  {journal} {Phys. Rev. E}\ }\textbf {\bibinfo {volume} {103}},\
  \bibinfo {pages} {L040101} (\bibinfo {year} {2021})}\BibitemShut {NoStop}%
\bibitem [{\citenamefont {Ball}\ \emph {et~al.}(2012)\citenamefont {Ball},
  \citenamefont {Connaughton}, \citenamefont {Jones}, \citenamefont {Rajesh},\
  and\ \citenamefont {Zaboronski}}]{ball2012collective}%
  \BibitemOpen
  \bibfield  {author} {\bibinfo {author} {\bibfnamefont {R.~C.}\ \bibnamefont
  {Ball}}, \bibinfo {author} {\bibfnamefont {C.}~\bibnamefont {Connaughton}},
  \bibinfo {author} {\bibfnamefont {P.~P.}\ \bibnamefont {Jones}}, \bibinfo
  {author} {\bibfnamefont {R.}~\bibnamefont {Rajesh}},\ and\ \bibinfo {author}
  {\bibfnamefont {O.}~\bibnamefont {Zaboronski}},\ }\bibfield  {title}
  {\bibinfo {title} {Collective oscillations in irreversible coagulation driven
  by monomer inputs and large-cluster outputs},\ }\href
  {https://doi.org/10.1103/PhysRevLett.109.168304} {\bibfield  {journal}
  {\bibinfo  {journal} {Phys. Rev. Lett.}\ }\textbf {\bibinfo {volume} {109}},\
  \bibinfo {pages} {168304} (\bibinfo {year} {2012})}\BibitemShut {NoStop}%
\bibitem [{\citenamefont {Matveev}\ \emph {et~al.}(2020)\citenamefont
  {Matveev}, \citenamefont {Sorokin}, \citenamefont {Smirnov},\ and\
  \citenamefont {Tyrtyshnikov}}]{matveev2020oscillating}%
  \BibitemOpen
  \bibfield  {author} {\bibinfo {author} {\bibfnamefont {S.~A.}\ \bibnamefont
  {Matveev}}, \bibinfo {author} {\bibfnamefont {A.~A.}\ \bibnamefont
  {Sorokin}}, \bibinfo {author} {\bibfnamefont {A.~P.}\ \bibnamefont
  {Smirnov}},\ and\ \bibinfo {author} {\bibfnamefont {E.~E.}\ \bibnamefont
  {Tyrtyshnikov}},\ }\bibfield  {title} {\bibinfo {title} {Oscillating
  stationary distributions of nanoclusters in an open system},\ }\href
  {https://doi.org/10.1080/13873954.2020.1793786} {\bibfield  {journal}
  {\bibinfo  {journal} {Mathematical and Computer Modelling of Dynamical
  Systems}\ }\textbf {\bibinfo {volume} {26}},\ \bibinfo {pages} {562}
  (\bibinfo {year} {2020})}\BibitemShut {NoStop}%
\bibitem [{\citenamefont {Ball}\ \emph {et~al.}(2011)\citenamefont {Ball},
  \citenamefont {Connaughton}, \citenamefont {Stein},\ and\ \citenamefont
  {Zaboronski}}]{ball2011instantaneous}%
  \BibitemOpen
  \bibfield  {author} {\bibinfo {author} {\bibfnamefont {R.~C.}\ \bibnamefont
  {Ball}}, \bibinfo {author} {\bibfnamefont {C.}~\bibnamefont {Connaughton}},
  \bibinfo {author} {\bibfnamefont {T.~H.}\ \bibnamefont {Stein}},\ and\
  \bibinfo {author} {\bibfnamefont {O.}~\bibnamefont {Zaboronski}},\ }\bibfield
   {title} {\bibinfo {title} {Instantaneous gelation in {S}moluchowski's
  coagulation equation revisited},\ }\href
  {https://doi.org/10.1103/PhysRevE.84.011111} {\bibfield  {journal} {\bibinfo
  {journal} {Phys. Rev. E}\ }\textbf {\bibinfo {volume} {84}},\ \bibinfo
  {pages} {011111} (\bibinfo {year} {2011})}\BibitemShut {NoStop}%
\bibitem [{\citenamefont {Field}\ and\ \citenamefont {Saslaw}(1965)}]{Saslaw}%
  \BibitemOpen
  \bibfield  {author} {\bibinfo {author} {\bibfnamefont {G.~B.}\ \bibnamefont
  {Field}}\ and\ \bibinfo {author} {\bibfnamefont {W.~C.}\ \bibnamefont
  {Saslaw}},\ }\bibfield  {title} {\bibinfo {title} {A statistical model of the
  formation of stars and interstellar clouds},\ }\href
  {https://adsabs.harvard.edu/pdf/1965ApJ...142..568F} {\bibfield  {journal}
  {\bibinfo  {journal} {Astrophys. J}\ }\textbf {\bibinfo {volume} {142}},\
  \bibinfo {pages} {568} (\bibinfo {year} {1965})}\BibitemShut {NoStop}%
\bibitem [{\citenamefont {White}(1982)}]{White82}%
  \BibitemOpen
  \bibfield  {author} {\bibinfo {author} {\bibfnamefont {W.~H.}\ \bibnamefont
  {White}},\ }\bibfield  {title} {\bibinfo {title} {On the form of steady-state
  solutions to the coagulation equations},\ }\href
  {https://www.sciencedirect.com/science/article/pii/0021979782903824}
  {\bibfield  {journal} {\bibinfo  {journal} {J. Colloid Interface Sci.}\
  }\textbf {\bibinfo {volume} {87}},\ \bibinfo {pages} {204} (\bibinfo {year}
  {1982})}\BibitemShut {NoStop}%
\bibitem [{\citenamefont {Krapivsky}\ and\ \citenamefont
  {Connaughton}(2012)}]{Colm12}%
  \BibitemOpen
  \bibfield  {author} {\bibinfo {author} {\bibfnamefont {P.~L.}\ \bibnamefont
  {Krapivsky}}\ and\ \bibinfo {author} {\bibfnamefont {C.}~\bibnamefont
  {Connaughton}},\ }\bibfield  {title} {\bibinfo {title} {Driven {B}rownian
  coagulation of polymers},\ }\href {https://doi.org/10.1063/1.4718833}
  {\bibfield  {journal} {\bibinfo  {journal} {J. Chem. Phys.}\ }\textbf
  {\bibinfo {volume} {136}} (\bibinfo {year} {2012})}\BibitemShut {NoStop}%
\bibitem [{\citenamefont {Fortin}\ and\ \citenamefont
  {Choi}(2023)}]{fortin2023stability}%
  \BibitemOpen
  \bibfield  {author} {\bibinfo {author} {\bibfnamefont {J.-Y.}\ \bibnamefont
  {Fortin}}\ and\ \bibinfo {author} {\bibfnamefont {M.}~\bibnamefont {Choi}},\
  }\bibfield  {title} {\bibinfo {title} {Stability condition of the steady
  oscillations in aggregation models with shattering process and
  self-fragmentation},\ }\href {https://doi.org/10.1088/1751-8121/acf3b9}
  {\bibfield  {journal} {\bibinfo  {journal} {J. Phys. A}\ }\textbf {\bibinfo
  {volume} {56}},\ \bibinfo {pages} {385004} (\bibinfo {year}
  {2023})}\BibitemShut {NoStop}%
\bibitem [{\citenamefont {Krapivsky}(1991)}]{krapivsky1991aggregation}%
  \BibitemOpen
  \bibfield  {author} {\bibinfo {author} {\bibfnamefont {P.~L.}\ \bibnamefont
  {Krapivsky}},\ }\bibfield  {title} {\bibinfo {title} {Aggregation processes
  with n-particle elementary reactions},\ }\href
  {https://doi.org/10.1088/0305-4470/24/19/028} {\bibfield  {journal} {\bibinfo
   {journal} {J. Phys. A}\ }\textbf {\bibinfo {volume} {24}},\ \bibinfo {pages}
  {4697} (\bibinfo {year} {1991})}\BibitemShut {NoStop}%
\bibitem [{\citenamefont {Jiang}(1994)}]{jiang1994scaling}%
  \BibitemOpen
  \bibfield  {author} {\bibinfo {author} {\bibfnamefont {Y.}~\bibnamefont
  {Jiang}},\ }\bibfield  {title} {\bibinfo {title} {Scaling theory for
  multipolymer coagulation},\ }\href {https://doi.org/10.1103/PhysRevE.50.4901}
  {\bibfield  {journal} {\bibinfo  {journal} {Phys. Rev. E}\ }\textbf {\bibinfo
  {volume} {50}},\ \bibinfo {pages} {4901} (\bibinfo {year}
  {1994})}\BibitemShut {NoStop}%
\bibitem [{\citenamefont {Krapivsky}(1994)}]{krapivsky1994diffusion}%
  \BibitemOpen
  \bibfield  {author} {\bibinfo {author} {\bibfnamefont {P.~L.}\ \bibnamefont
  {Krapivsky}},\ }\bibfield  {title} {\bibinfo {title}
  {Diffusion-limited-aggregation processes with three-particle elementary
  reactions},\ }\href {https://doi.org/10.1103/PhysRevE.49.3233} {\bibfield
  {journal} {\bibinfo  {journal} {Phys. Rev. E}\ }\textbf {\bibinfo {volume}
  {49}},\ \bibinfo {pages} {3233} (\bibinfo {year} {1994})}\BibitemShut
  {NoStop}%
\bibitem [{\citenamefont {Oshanin}\ \emph {et~al.}(1995)\citenamefont
  {Oshanin}, \citenamefont {Stemmer}, \citenamefont {Luding},\ and\
  \citenamefont {Blumen}}]{oshanin1995smoluchowski}%
  \BibitemOpen
  \bibfield  {author} {\bibinfo {author} {\bibfnamefont {G.}~\bibnamefont
  {Oshanin}}, \bibinfo {author} {\bibfnamefont {A.}~\bibnamefont {Stemmer}},
  \bibinfo {author} {\bibfnamefont {S.}~\bibnamefont {Luding}},\ and\ \bibinfo
  {author} {\bibfnamefont {A.}~\bibnamefont {Blumen}},\ }\bibfield  {title}
  {\bibinfo {title} {Smoluchowski approach for three-body reactions in one
  dimension},\ }\href {https://doi.org/10.1103/PhysRevE.52.5800} {\bibfield
  {journal} {\bibinfo  {journal} {Phys. Rev. E}\ }\textbf {\bibinfo {volume}
  {52}},\ \bibinfo {pages} {5800} (\bibinfo {year} {1995})}\BibitemShut
  {NoStop}%
\bibitem [{\citenamefont {Brener}(2014)}]{brener2014model}%
  \BibitemOpen
  \bibfield  {author} {\bibinfo {author} {\bibfnamefont {A.~M.}\ \bibnamefont
  {Brener}},\ }\bibfield  {title} {\bibinfo {title} {Model of many-particle
  aggregation in dense particle systems},\ }\href
  {https://doi.org/10.3303/CET1438025} {\bibfield  {journal} {\bibinfo
  {journal} {Chem. Eng.}\ }\textbf {\bibinfo {volume} {38}},\ \bibinfo {pages}
  {145} (\bibinfo {year} {2014})}\BibitemShut {NoStop}%
\bibitem [{\citenamefont {Jiang}(1996)}]{jiang1996instantaneous}%
  \BibitemOpen
  \bibfield  {author} {\bibinfo {author} {\bibfnamefont {Y.}~\bibnamefont
  {Jiang}},\ }\bibfield  {title} {\bibinfo {title} {Instantaneous gelation in
  the generalized {S}moluchovski coagulation equation},\ }\href
  {https://doi.org/10.1088/0305-4470/29/24/014} {\bibfield  {journal} {\bibinfo
   {journal} {J. Phys. A}\ }\textbf {\bibinfo {volume} {29}},\ \bibinfo {pages}
  {7893} (\bibinfo {year} {1996})}\BibitemShut {NoStop}%
\bibitem [{\citenamefont {Lushnikov}\ and\ \citenamefont
  {Piskunov}(1983)}]{lushnikov1983analytic}%
  \BibitemOpen
  \bibfield  {author} {\bibinfo {author} {\bibfnamefont {A.~A.}\ \bibnamefont
  {Lushnikov}}\ and\ \bibinfo {author} {\bibfnamefont {V.~N.}\ \bibnamefont
  {Piskunov}},\ }\bibfield  {title} {\bibinfo {title} {Analytic solutions in
  the theory of coagulating systems with sinks},\ }\href
  {https://doi.org/10.1016/0021-8928(83)90110-7} {\bibfield  {journal}
  {\bibinfo  {journal} {J. Appl. Math. Mech.}\ }\textbf {\bibinfo {volume}
  {47}},\ \bibinfo {pages} {743} (\bibinfo {year} {1983})}\BibitemShut
  {NoStop}%
\bibitem [{\citenamefont {Erd{\H{o}}s}\ and\ \citenamefont
  {R{\'e}nyi}(1960)}]{ER60}%
  \BibitemOpen
  \bibfield  {author} {\bibinfo {author} {\bibfnamefont {P.}~\bibnamefont
  {Erd{\H{o}}s}}\ and\ \bibinfo {author} {\bibfnamefont {A.}~\bibnamefont
  {R{\'e}nyi}},\ }\bibfield  {title} {\bibinfo {title} {On the evolution of
  random graphs},\ }\href@noop {} {\bibfield  {journal} {\bibinfo  {journal}
  {Publ. Math. Inst.}\ }\textbf {\bibinfo {volume} {5}},\ \bibinfo {pages} {17}
  (\bibinfo {year} {1960})}\BibitemShut {NoStop}%
\bibitem [{\citenamefont {Janson}\ \emph {et~al.}(1993)\citenamefont {Janson},
  \citenamefont {Knuth}, \citenamefont {{\L}uczak},\ and\ \citenamefont
  {Pittel}}]{Janson93}%
  \BibitemOpen
  \bibfield  {author} {\bibinfo {author} {\bibfnamefont {S.}~\bibnamefont
  {Janson}}, \bibinfo {author} {\bibfnamefont {D.~E.}\ \bibnamefont {Knuth}},
  \bibinfo {author} {\bibfnamefont {T.}~\bibnamefont {{\L}uczak}},\ and\
  \bibinfo {author} {\bibfnamefont {B.}~\bibnamefont {Pittel}},\ }\bibfield
  {title} {\bibinfo {title} {The birth of the giant component},\ }\href
  {https://doi.org/10.1002/rsa.3240040303} {\bibfield  {journal} {\bibinfo
  {journal} {Random Structures \& Algorithms}\ }\textbf {\bibinfo {volume}
  {4}},\ \bibinfo {pages} {233} (\bibinfo {year} {1993})}\BibitemShut {NoStop}%
\bibitem [{\citenamefont {Smoluchowski}(1918)}]{Smol17}%
  \BibitemOpen
  \bibfield  {author} {\bibinfo {author} {\bibfnamefont {M.~v.}\ \bibnamefont
  {Smoluchowski}},\ }\bibfield  {title} {\bibinfo {title} {Versuch einer
  mathematischen {T}heorie der {K}oagulationskinetik kolloider
  {L}{\"o}sungen},\ }\href {https://doi.org/10.1515/zpch-1918-9209} {\bibfield
  {journal} {\bibinfo  {journal} {Zeit. Physikalische Chemie}\ }\textbf
  {\bibinfo {volume} {92}},\ \bibinfo {pages} {129} (\bibinfo {year}
  {1918})}\BibitemShut {NoStop}%
\bibitem [{\citenamefont {Chandrasekhar}(1943)}]{Chandra43}%
  \BibitemOpen
  \bibfield  {author} {\bibinfo {author} {\bibfnamefont {S.}~\bibnamefont
  {Chandrasekhar}},\ }\bibfield  {title} {\bibinfo {title} {Stochastic problems
  in physics and astronomy},\ }\href {https://doi.org/10.1103/RevModPhys.15.1}
  {\bibfield  {journal} {\bibinfo  {journal} {Reviews of modern physics}\
  }\textbf {\bibinfo {volume} {15}},\ \bibinfo {pages} {1} (\bibinfo {year}
  {1943})}\BibitemShut {NoStop}%
\bibitem [{\citenamefont {Stefonishin}\ \emph {et~al.}(2019)\citenamefont
  {Stefonishin}, \citenamefont {Matveev},\ and\ \citenamefont
  {Zheltkov}}]{stefonishin2019tensors}%
  \BibitemOpen
  \bibfield  {author} {\bibinfo {author} {\bibfnamefont {D.~A.}\ \bibnamefont
  {Stefonishin}}, \bibinfo {author} {\bibfnamefont {S.~A.}\ \bibnamefont
  {Matveev}},\ and\ \bibinfo {author} {\bibfnamefont {D.~A.}\ \bibnamefont
  {Zheltkov}},\ }\bibfield  {title} {\bibinfo {title} {Tensors in modelling
  multi-particle interactions},\ }in\ \href
  {https://doi.org/10.1007/978-3-030-41032-2_19} {\emph {\bibinfo {booktitle}
  {International Conference on Large-Scale Scientific Computing}}},\ \bibinfo
  {editor} {edited by\ \bibinfo {editor} {\bibfnamefont {I.}~\bibnamefont
  {Lirkov}}\ and\ \bibinfo {editor} {\bibfnamefont {S.}~\bibnamefont
  {Margenov}}}\ (\bibinfo {organization} {Springer, Cham},\ \bibinfo {year}
  {2019})\ pp.\ \bibinfo {pages} {173--180}\BibitemShut {NoStop}%
\bibitem [{\citenamefont {Lukashevich}\ \emph {et~al.}(2022)\citenamefont
  {Lukashevich}, \citenamefont {Ovchinnikov}, \citenamefont {Tyukin},
  \citenamefont {Matveev},\ and\ \citenamefont
  {Brilliantov}}]{lukashevich2022data}%
  \BibitemOpen
  \bibfield  {author} {\bibinfo {author} {\bibfnamefont {D.}~\bibnamefont
  {Lukashevich}}, \bibinfo {author} {\bibfnamefont {G.}~\bibnamefont
  {Ovchinnikov}}, \bibinfo {author} {\bibfnamefont {I.~Y.}\ \bibnamefont
  {Tyukin}}, \bibinfo {author} {\bibfnamefont {S.~A.}\ \bibnamefont
  {Matveev}},\ and\ \bibinfo {author} {\bibfnamefont {N.~V.}\ \bibnamefont
  {Brilliantov}},\ }\bibfield  {title} {\bibinfo {title} {Data-driven approach
  for modeling coagulation kinetics},\ }\href
  {https://doi.org/10.1007/s10598-023-09574-5} {\bibfield  {journal} {\bibinfo
  {journal} {Computational Mathematics and Modeling}\ }\textbf {\bibinfo
  {volume} {33}},\ \bibinfo {pages} {310} (\bibinfo {year} {2022})}\BibitemShut
  {NoStop}%
\bibitem [{\citenamefont {Oseledets}(2011)}]{oseledets2011tensor}%
  \BibitemOpen
  \bibfield  {author} {\bibinfo {author} {\bibfnamefont {I.~V.}\ \bibnamefont
  {Oseledets}},\ }\bibfield  {title} {\bibinfo {title} {Tensor-train
  decomposition},\ }\href {https://doi.org/10.1137/090752286} {\bibfield
  {journal} {\bibinfo  {journal} {SIAM J. Sci. Comput.}\ }\textbf {\bibinfo
  {volume} {33}},\ \bibinfo {pages} {2295} (\bibinfo {year}
  {2011})}\BibitemShut {NoStop}%
\bibitem [{\citenamefont {Bro}(1997)}]{bro1997parafac}%
  \BibitemOpen
  \bibfield  {author} {\bibinfo {author} {\bibfnamefont {R.}~\bibnamefont
  {Bro}},\ }\bibfield  {title} {\bibinfo {title} {{P}{A}{R}{A}{F}{A}{C}.
  {T}utorial and applications},\ }\href
  {https://doi.org/10.1016/S0169-7439(97)00032-4} {\bibfield  {journal}
  {\bibinfo  {journal} {Chemometrics and intelligent laboratory systems}\
  }\textbf {\bibinfo {volume} {38}},\ \bibinfo {pages} {149} (\bibinfo {year}
  {1997})}\BibitemShut {NoStop}%
\bibitem [{\citenamefont {Flory}(1941{\natexlab{a}})}]{Flory41}%
  \BibitemOpen
  \bibfield  {author} {\bibinfo {author} {\bibfnamefont {P.~J.}\ \bibnamefont
  {Flory}},\ }\bibfield  {title} {\bibinfo {title} {Molecular size distribution
  in three dimensional polymers. {I}. {G}elation},\ }\href
  {https://doi.org/10.1021/ja01856a061} {\bibfield  {journal} {\bibinfo
  {journal} {J. Amer. Chem. Soc.}\ }\textbf {\bibinfo {volume} {63}},\ \bibinfo
  {pages} {3083} (\bibinfo {year} {1941}{\natexlab{a}})}\BibitemShut {NoStop}%
\bibitem [{\citenamefont {Flory}(1941{\natexlab{b}})}]{Flory41b}%
  \BibitemOpen
  \bibfield  {author} {\bibinfo {author} {\bibfnamefont {P.~J.}\ \bibnamefont
  {Flory}},\ }\bibfield  {title} {\bibinfo {title} {Molecular size distribution
  in three dimensional polymers. {II}. {T}rifunctional branching units},\
  }\href {https://doi.org/10.1021/ja01856a062} {\bibfield  {journal} {\bibinfo
  {journal} {J. Amer. Chem. Soc.}\ }\textbf {\bibinfo {volume} {63}},\ \bibinfo
  {pages} {3091} (\bibinfo {year} {1941}{\natexlab{b}})}\BibitemShut {NoStop}%
\bibitem [{\citenamefont {Flory}(1941{\natexlab{c}})}]{Flory41c}%
  \BibitemOpen
  \bibfield  {author} {\bibinfo {author} {\bibfnamefont {P.~J.}\ \bibnamefont
  {Flory}},\ }\bibfield  {title} {\bibinfo {title} {Molecular size distribution
  in three dimensional polymers. {III}. {T}etrafunctional branching units},\
  }\href {https://doi.org/10.1021/ja01856a063} {\bibfield  {journal} {\bibinfo
  {journal} {J. Amer. Chem. Soc.}\ }\textbf {\bibinfo {volume} {63}},\ \bibinfo
  {pages} {3096} (\bibinfo {year} {1941}{\natexlab{c}})}\BibitemShut {NoStop}%
\bibitem [{\citenamefont {Stockmayer}(1943)}]{Stockmayer43}%
  \BibitemOpen
  \bibfield  {author} {\bibinfo {author} {\bibfnamefont {W.~H.}\ \bibnamefont
  {Stockmayer}},\ }\bibfield  {title} {\bibinfo {title} {Theory of molecular
  size distribution and gel formation in branched-chain polymers},\ }\href
  {https://doi.org/10.1063/1.1723803} {\bibfield  {journal} {\bibinfo
  {journal} {J. Chem. Phys.}\ }\textbf {\bibinfo {volume} {11}},\ \bibinfo
  {pages} {45} (\bibinfo {year} {1943})}\BibitemShut {NoStop}%
\bibitem [{\citenamefont {Lushnikov}(2004)}]{Lushnikov04}%
  \BibitemOpen
  \bibfield  {author} {\bibinfo {author} {\bibfnamefont {A.~A.}\ \bibnamefont
  {Lushnikov}},\ }\bibfield  {title} {\bibinfo {title} {From sol to gel
  exactly},\ }\href {https://doi.org/10.1103/PhysRevLett.93.198302} {\bibfield
  {journal} {\bibinfo  {journal} {Phys. Rev. Lett.}\ }\textbf {\bibinfo
  {volume} {93}},\ \bibinfo {pages} {198302} (\bibinfo {year}
  {2004})}\BibitemShut {NoStop}%
\bibitem [{\citenamefont {Lushnikov}(2005)}]{Lushnikov05}%
  \BibitemOpen
  \bibfield  {author} {\bibinfo {author} {\bibfnamefont {A.~A.}\ \bibnamefont
  {Lushnikov}},\ }\bibfield  {title} {\bibinfo {title} {Exact kinetics of the
  sol-gel transition},\ }\href {https://doi.org/10.1103/PhysRevE.71.046129}
  {\bibfield  {journal} {\bibinfo  {journal} {Phys. Rev. E}\ }\textbf {\bibinfo
  {volume} {71}},\ \bibinfo {pages} {046129} (\bibinfo {year}
  {2005})}\BibitemShut {NoStop}%
\bibitem [{\citenamefont {Ben-Naim}\ and\ \citenamefont
  {Krapivsky}(2005)}]{BN-RG}%
  \BibitemOpen
  \bibfield  {author} {\bibinfo {author} {\bibfnamefont {E.}~\bibnamefont
  {Ben-Naim}}\ and\ \bibinfo {author} {\bibfnamefont {P.~L.}\ \bibnamefont
  {Krapivsky}},\ }\bibfield  {title} {\bibinfo {title} {Kinetic theory of
  random graphs: From paths to cycles},\ }\href
  {https://doi.org/10.1103/PhysRevE.71.026129} {\bibfield  {journal} {\bibinfo
  {journal} {Phys. Rev. E}\ }\textbf {\bibinfo {volume} {71}},\ \bibinfo
  {pages} {026129} (\bibinfo {year} {2005})}\BibitemShut {NoStop}%
\bibitem [{\citenamefont {Leyvraz}(2022)}]{Leyvraz22}%
  \BibitemOpen
  \bibfield  {author} {\bibinfo {author} {\bibfnamefont {F.}~\bibnamefont
  {Leyvraz}},\ }\bibfield  {title} {\bibinfo {title} {Rate equation limit for a
  combinatorial solution of a stochastic aggregation model},\ }\href
  {https://doi.org/10.1103/PhysRevE.106.024133} {\bibfield  {journal} {\bibinfo
   {journal} {Phys. Rev. E}\ }\textbf {\bibinfo {volume} {106}},\ \bibinfo
  {pages} {024133} (\bibinfo {year} {2022})}\BibitemShut {NoStop}%
\bibitem [{\citenamefont {Bollob\'{a}s}(2001)}]{Bollobas}%
  \BibitemOpen
  \bibfield  {author} {\bibinfo {author} {\bibfnamefont {B.}~\bibnamefont
  {Bollob\'{a}s}},\ }\href {https://doi.org/10.1017/CBO9780511814068} {\emph
  {\bibinfo {title} {Random {G}raphs}}}\ (\bibinfo  {publisher} {Cambridge
  University Press},\ \bibinfo {address} {Cambridge, UK},\ \bibinfo {year}
  {2001})\BibitemShut {NoStop}%
\bibitem [{\citenamefont {Flajolet}\ and\ \citenamefont
  {Sedgewick}(2009)}]{Flajolet}%
  \BibitemOpen
  \bibfield  {author} {\bibinfo {author} {\bibfnamefont {P.}~\bibnamefont
  {Flajolet}}\ and\ \bibinfo {author} {\bibfnamefont {R.}~\bibnamefont
  {Sedgewick}},\ }\href {https://doi.org/10.1017/CBO9780511801655} {\emph
  {\bibinfo {title} {Analytic combinatorics}}}\ (\bibinfo  {publisher}
  {Cambridge University Press},\ \bibinfo {year} {2009})\BibitemShut {NoStop}%
\bibitem [{\citenamefont {Newman}(2010)}]{Newman-book}%
  \BibitemOpen
  \bibfield  {author} {\bibinfo {author} {\bibfnamefont {M.}~\bibnamefont
  {Newman}},\ }\href
  {https://doi.org/10.1093/acprof:oso/9780199206650.001.0001} {\emph {\bibinfo
  {title} {Networks: {A}n {I}ntroduction}}}\ (\bibinfo  {publisher} {Oxford
  University Press},\ \bibinfo {address} {New York, USA},\ \bibinfo {year}
  {2010})\BibitemShut {NoStop}%
\bibitem [{\citenamefont {van~der Hofstad}(2016)}]{Hofstad}%
  \BibitemOpen
  \bibfield  {author} {\bibinfo {author} {\bibfnamefont {R.}~\bibnamefont
  {van~der Hofstad}},\ }\href {https://doi.org/10.1017/9781316779422} {\emph
  {\bibinfo {title} {Random graphs and complex networks}}}\ (\bibinfo
  {publisher} {Cambridge University Press},\ \bibinfo {address} {Cambridge,
  UK},\ \bibinfo {year} {2016})\BibitemShut {NoStop}%
\bibitem [{\citenamefont {Frieze}\ and\ \citenamefont
  {Karo\'{n}ski}(2016)}]{Frieze}%
  \BibitemOpen
  \bibfield  {author} {\bibinfo {author} {\bibfnamefont {A.}~\bibnamefont
  {Frieze}}\ and\ \bibinfo {author} {\bibfnamefont {M.}~\bibnamefont
  {Karo\'{n}ski}},\ }\href {https://doi.org/10.1017/CBO9781316339831} {\emph
  {\bibinfo {title} {Introduction to {R}andom {G}raphs}}}\ (\bibinfo
  {publisher} {Cambridge University Press},\ \bibinfo {address} {Cambridge,
  UK},\ \bibinfo {year} {2016})\BibitemShut {NoStop}%
\bibitem [{\citenamefont {Chatterjee}(2016)}]{chatterjee16}%
  \BibitemOpen
  \bibfield  {author} {\bibinfo {author} {\bibfnamefont {S.}~\bibnamefont
  {Chatterjee}},\ }\bibfield  {title} {\bibinfo {title} {An introduction to
  large deviations for random graphs},\ }\href
  {https://doi.org/10.1090/bull/1539} {\bibfield  {journal} {\bibinfo
  {journal} {Bull. Amer. Math. Soc.}\ }\textbf {\bibinfo {volume} {53}},\
  \bibinfo {pages} {617} (\bibinfo {year} {2016})}\BibitemShut {NoStop}%
\bibitem [{\citenamefont {Ziff}\ and\ \citenamefont {Stell}(1980)}]{Ziff80}%
  \BibitemOpen
  \bibfield  {author} {\bibinfo {author} {\bibfnamefont {R.~M.}\ \bibnamefont
  {Ziff}}\ and\ \bibinfo {author} {\bibfnamefont {G.}~\bibnamefont {Stell}},\
  }\bibfield  {title} {\bibinfo {title} {Kinetics of polymer gelation},\ }\href
  {https://doi.org/10.1063/1.440502} {\bibfield  {journal} {\bibinfo  {journal}
  {J. Chem. Phys.}\ }\textbf {\bibinfo {volume} {73}},\ \bibinfo {pages} {3492}
  (\bibinfo {year} {1980})}\BibitemShut {NoStop}%
\bibitem [{\citenamefont {Tanaka}\ and\ \citenamefont
  {Ishida}(1994)}]{tanaka1994}%
  \BibitemOpen
  \bibfield  {author} {\bibinfo {author} {\bibfnamefont {F.}~\bibnamefont
  {Tanaka}}\ and\ \bibinfo {author} {\bibfnamefont {M.}~\bibnamefont
  {Ishida}},\ }\bibfield  {title} {\bibinfo {title} {Phase formation of
  two-component physical gels},\ }\href
  {https://www.sciencedirect.com/science/article/pii/0378437194904537}
  {\bibfield  {journal} {\bibinfo  {journal} {Physica A}\ }\textbf {\bibinfo
  {volume} {204}},\ \bibinfo {pages} {660} (\bibinfo {year}
  {1994})}\BibitemShut {NoStop}%
\bibitem [{\citenamefont {Tanaka}(2011)}]{tanaka2011book}%
  \BibitemOpen
  \bibfield  {author} {\bibinfo {author} {\bibfnamefont {F.}~\bibnamefont
  {Tanaka}},\ }\href {https://doi.org/10.1017/CBO9780511975691} {\emph
  {\bibinfo {title} {Polymer physics: applications to molecular association and
  thermoreversible gelation}}}\ (\bibinfo  {publisher} {Cambridge University
  Press},\ \bibinfo {address} {Cambridge, UK},\ \bibinfo {year}
  {2011})\BibitemShut {NoStop}%
\bibitem [{\citenamefont {Graham}\ \emph {et~al.}(1994)\citenamefont {Graham},
  \citenamefont {Knuth},\ and\ \citenamefont {Patashnik}}]{Knuth}%
  \BibitemOpen
  \bibfield  {author} {\bibinfo {author} {\bibfnamefont {R.~L.}\ \bibnamefont
  {Graham}}, \bibinfo {author} {\bibfnamefont {D.~E.}\ \bibnamefont {Knuth}},\
  and\ \bibinfo {author} {\bibfnamefont {O.}~\bibnamefont {Patashnik}},\
  }\href@noop {} {\emph {\bibinfo {title} {Concrete {M}athematics: {A}
  {F}oundation for {C}omputer {S}cience}}}\ (\bibinfo  {publisher}
  {Addison-Wesley},\ \bibinfo {address} {Reading, Massachusetts},\ \bibinfo
  {year} {1994})\BibitemShut {NoStop}%
\bibitem [{\citenamefont {Thamm}\ and\ \citenamefont
  {Erukhimovich}(2003)}]{thamm2003phase}%
  \BibitemOpen
  \bibfield  {author} {\bibinfo {author} {\bibfnamefont {M.~V.}\ \bibnamefont
  {Thamm}}\ and\ \bibinfo {author} {\bibfnamefont {I.~Y.}\ \bibnamefont
  {Erukhimovich}},\ }\bibfield  {title} {\bibinfo {title} {Phase diagrams
  classification of the systems with thermoreversible alternating association
  (the {F}lory approach)},\ }\href {https://doi.org/10.1063/1.1586253}
  {\bibfield  {journal} {\bibinfo  {journal} {J. Chem. Phys.}\ }\textbf
  {\bibinfo {volume} {119}},\ \bibinfo {pages} {2720} (\bibinfo {year}
  {2003})}\BibitemShut {NoStop}%
\bibitem [{\citenamefont {Lushnikov}(2015)}]{lushnikov2015source}%
  \BibitemOpen
  \bibfield  {author} {\bibinfo {author} {\bibfnamefont {A.~A.}\ \bibnamefont
  {Lushnikov}},\ }\bibfield  {title} {\bibinfo {title} {Source-enhanced
  coalescence of trees in a random forest},\ }\href
  {https://doi.org/10.1103/PhysRevE.92.022135} {\bibfield  {journal} {\bibinfo
  {journal} {Phys. Rev. E}\ }\textbf {\bibinfo {volume} {92}},\ \bibinfo
  {pages} {022135} (\bibinfo {year} {2015})}\BibitemShut {NoStop}%
\bibitem [{\citenamefont {Contat}\ and\ \citenamefont
  {Curien}(2023)}]{Curien23}%
  \BibitemOpen
  \bibfield  {author} {\bibinfo {author} {\bibfnamefont {A.}~\bibnamefont
  {Contat}}\ and\ \bibinfo {author} {\bibfnamefont {N.}~\bibnamefont
  {Curien}},\ }\bibfield  {title} {\bibinfo {title} {Parking on {C}ayley trees
  and frozen {Erd\H{os}-R\'{e}nyi}},\ }\href
  {https://doi.org/10.1214/23-AOP1632} {\bibfield  {journal} {\bibinfo
  {journal} {Ann. Probab.}\ }\textbf {\bibinfo {volume} {51}},\ \bibinfo
  {pages} {1993} (\bibinfo {year} {2023})}\BibitemShut {NoStop}%
\bibitem [{\citenamefont {Krapivsky}(2023)}]{K24-SRG}%
  \BibitemOpen
  \bibfield  {author} {\bibinfo {author} {\bibfnamefont {P.~L.}\ \bibnamefont
  {Krapivsky}},\ }\bibfield  {title} {\bibinfo {title} {Simple evolving random
  graphs},\ }\href {https://arxiv.org/abs/2312.02952} {\bibfield  {journal}
  {\bibinfo  {journal} {arXiv:2312.02952}\ } (\bibinfo {year}
  {2023})}\BibitemShut {NoStop}%
\bibitem [{\citenamefont {Oseledets}\ and\ \citenamefont
  {Tyrtyshnikov}(2010)}]{oseledets2010tt}%
  \BibitemOpen
  \bibfield  {author} {\bibinfo {author} {\bibfnamefont {I.}~\bibnamefont
  {Oseledets}}\ and\ \bibinfo {author} {\bibfnamefont {E.}~\bibnamefont
  {Tyrtyshnikov}},\ }\bibfield  {title} {\bibinfo {title} {{TT}-cross
  approximation for multidimensional arrays},\ }\href
  {https://doi.org/10.1016/j.laa.2009.07.024} {\bibfield  {journal} {\bibinfo
  {journal} {Linear Alg. Appl.}\ }\textbf {\bibinfo {volume} {432}},\ \bibinfo
  {pages} {70} (\bibinfo {year} {2010})}\BibitemShut {NoStop}%
\bibitem [{\citenamefont {Cooley}\ and\ \citenamefont
  {Tukey}(1965)}]{cooley1965algorithm}%
  \BibitemOpen
  \bibfield  {author} {\bibinfo {author} {\bibfnamefont {J.~W.}\ \bibnamefont
  {Cooley}}\ and\ \bibinfo {author} {\bibfnamefont {J.~W.}\ \bibnamefont
  {Tukey}},\ }\bibfield  {title} {\bibinfo {title} {An algorithm for the
  machine calculation of complex {F}ourier series},\ }\href
  {https://doi.org/10.2307/2003354} {\bibfield  {journal} {\bibinfo  {journal}
  {Mathematics of computation}\ }\textbf {\bibinfo {volume} {19}},\ \bibinfo
  {pages} {297} (\bibinfo {year} {1965})}\BibitemShut {NoStop}%
\bibitem [{\citenamefont {Lee}(2000)}]{lee2000validity}%
  \BibitemOpen
  \bibfield  {author} {\bibinfo {author} {\bibfnamefont {M.~H.}\ \bibnamefont
  {Lee}},\ }\bibfield  {title} {\bibinfo {title} {On the validity of the
  coagulation equation and the nature of runaway growth},\ }\href
  {https://doi.org/10.1006/icar.1999.6239} {\bibfield  {journal} {\bibinfo
  {journal} {Icarus}\ }\textbf {\bibinfo {volume} {143}},\ \bibinfo {pages}
  {74} (\bibinfo {year} {2000})}\BibitemShut {NoStop}%
\bibitem [{\citenamefont {Dandekar}\ \emph {et~al.}(2023)\citenamefont
  {Dandekar}, \citenamefont {Rajesh}, \citenamefont {Subashri},\ and\
  \citenamefont {Zaboronski}}]{dandekar2023monte}%
  \BibitemOpen
  \bibfield  {author} {\bibinfo {author} {\bibfnamefont {R.}~\bibnamefont
  {Dandekar}}, \bibinfo {author} {\bibfnamefont {R.}~\bibnamefont {Rajesh}},
  \bibinfo {author} {\bibfnamefont {V.}~\bibnamefont {Subashri}},\ and\
  \bibinfo {author} {\bibfnamefont {O.}~\bibnamefont {Zaboronski}},\ }\bibfield
   {title} {\bibinfo {title} {A {M}onte {C}arlo algorithm to measure
  probabilities of rare events in cluster-cluster aggregation},\ }\href
  {https://doi.org/10.1016/j.cpc.2023.108727} {\bibfield  {journal} {\bibinfo
  {journal} {Computer Physics Communications}\ }\textbf {\bibinfo {volume}
  {288}},\ \bibinfo {pages} {108727} (\bibinfo {year} {2023})}\BibitemShut
  {NoStop}%
\bibitem [{\citenamefont {Dyachenko}\ \emph {et~al.}(2023)\citenamefont
  {Dyachenko}, \citenamefont {Matveev},\ and\ \citenamefont
  {Krapivsky}}]{dyachenko2023finite}%
  \BibitemOpen
  \bibfield  {author} {\bibinfo {author} {\bibfnamefont {R.~R.}\ \bibnamefont
  {Dyachenko}}, \bibinfo {author} {\bibfnamefont {S.~A.}\ \bibnamefont
  {Matveev}},\ and\ \bibinfo {author} {\bibfnamefont {P.~L.}\ \bibnamefont
  {Krapivsky}},\ }\bibfield  {title} {\bibinfo {title} {Finite-size effects in
  addition and chipping processes},\ }\href
  {https://doi.org/10.1103/PhysRevE.108.044119} {\bibfield  {journal} {\bibinfo
   {journal} {Phys. Rev. E}\ }\textbf {\bibinfo {volume} {108}},\ \bibinfo
  {pages} {044119} (\bibinfo {year} {2023})}\BibitemShut {NoStop}%
\end{thebibliography}%

\end{document}